\documentclass[twocolumn,aps,pra,showpacs]{revtex4}

\usepackage{epsfig}
\usepackage{graphics}
\usepackage{amssymb}

\newcommand{\bfr}{{\bf r}}

\newcommand{\bfk}{{\bf k}}

\newcommand{\bfx}{{\bf x}}
\newcommand{\bfy}{{\bf y}}
\newcommand{\bfz}{{\bf z}}
\newcommand{\bfd}{{\bf d}}
\newcommand{\bfe}{{\bf e}}
\newcommand{\bfE}{{\bf E}}
\newcommand{\bfP}{{\bf P}}

\newcommand{\tpsi}{\tilde{\psi}}
\newcommand{\tG}{\tilde{G}}
\newcommand{\tOmega}{\tilde{\Omega}}
\newcommand{\tDelta}{\tilde{\Delta}}
\newcommand{\tgamma}{\tilde{\gamma}}

\newcommand{\ddt}{\partial_t}
\newcommand{\ppt}{\partial_t}

\newcommand{\pptau}{\partial_\tau}

\newcommand{\ddtau}{\partial_\tau}
\newcommand{\nn}{\nonumber \\}
\newcommand{\beq}{\begin{equation}}
\newcommand{\eeq}{\end{equation}}
\newcommand{\beqa}{\begin{eqnarray}}
\newcommand{\eeqa}{\end{eqnarray}}
\newcommand{\la}{\langle}
\newcommand{\ra}{\rangle}
\newcommand{\vk}{\vec{k}}
\renewcommand{\vr}{\vec{r}}
\newcommand{\cF}{{\cal F}}
\newcommand{\cA}{{\cal A}}
\newcommand{\str}{^\ast}

\begin{document}


\title{Resonant Matter Wave Amplification in Mean Field Theory}



\author{T. C. Doan}
\author{Y. P. Huang}
\author{S. F. Wolf}
\author{M. G. Moore}
\affiliation{Department of Physics and Astronomy, Michigan State
University, East Lansing, MI 48824}


\date{\today}

\begin{abstract}
We develop a Green's function based mean-field theory for coherent mixing of matter- and light-waves. To demonstrate the 
utility of this approach, we analyse a co-propagating Raman matter-wave amplifier.
We find that for a given laser intensity, a significantly faster amplification
process can be achieved employing resonant rather than
off-resonance driving. The ratio of the matter-wave gain to atom loss-rate due to spontaneous emission is given by the optical depth of the sample, and is the same both on- and off-resonance.
Furthermore, we show that for short-times, the single-mode approximation for the matter-waves gives exact agreement with the full spatial dynamics. For long times, the off-resonant case shows suppressed amplification due to a spatially inhomogenous AC Stark shift associated with laser depletion. This suppression is absent on-resonance, where the AC Stark shift is absent.
\end{abstract}

\maketitle
\section{Introduction}
\label{intro}

Matter-wave amplification (MWA) in atomic ensembles is a stimulated (i.e. collectively enhanced)
nonlinear wave-mixing process in which two coherent matter waves
(pump 1 + signal) and one coherent light wave (pump 2) are input,
resulting in amplification of the signal matter wave, along with the phase-coherent
generation of an additional light wave (idler), whose wave-vector is
determined by phase-matching conditions \cite{LawBig98,MooMey99a,InoChiSta99,
KozSuzTor99,InoPfaGup99, MooMey99b,KruBurAud99,
MusYou00,InoLowGup00,Tri01, SchTorBoy03,PioBerRob03,
SchGreEri04,YosSugTor04, SchCamStr04, AveTri04,
ColPio04,Tri05,BonPioRob05, HilKamTar08}. As the amplification
process preserves the relative phase between the two matter waves,
MWA has been viewed as an active device with potential applications
in matter-wave interferometry
\cite{RolPhi02,SchHofAnd05,Lee06,PezSme06,JoShiWil07,HuaMoo08}. This
point of view, however, is somewhat limited by the fact that the signal is
amplified at the expense of pump 1, whose depletion eventually
suppresses the MWA gain mechanism. On the other hand, when viewed as
a fully quantum-mechanical process, a MWA-tyle process can in principle
map an arbitrary quantum state from a two-mode atomic
field (pump 1 + signal) onto a two-mode optical field (pump 2 +
idler) \cite{KozSuzTor99,InoChiSta99,SchCamStr04}. Thus quantum information
acquired by a matter-wave field could, e.g., be transferred to an optical field for
detection or long-distance transfer to a second bimodal atomic
system \cite{SheKraOls06,GinGarVes07}.

As a mechanism for information transfer, MWA is complementary to electromagnetically-induced-transparancy (EIT) based quantum information retrieval techniques \cite{FleImaMar05,DuaMon08}. Whereas EIT approaches use an adiabatic `counter-intuitive' pulse sequence to deplete the small signal mode, while mapping the stored information onto an outgoing light pulse; the MWA process uses a dynamical `intuitive' pulse sequence to amplify the small signal state, mapping a complementary quantum state, corresponding to the amplified signal state minus the original signal state, onto the outgoing light pulse. For the idealized case of 100\% pump-to-signal conversion, this complementary state corresponds to the quantum state of pump 1. If the initial populations of the pump and signal matter-waves are equal, then the EIT and MWA information retrieval techniques would be effectively indistinguishable.

In developing quantum information
protocols based on atomic systems, a universal challenge  is  to overcome the inevitable collisional dephasing and/or
environment-induced decoherence processes. For MWA-based devices, one possible solution is to use very strong pumping fields to complete the process  on a time-scale
where dephasing/decoherence is negligible. In order to determine if this approach is feasible, we must first
determine the ultimate rate limit to MWA.  Most MWA and closely-related `matter-wave superradiance' experiments  \cite{InoPfaGup99,KozSuzTor99,SchTorBoy03,SchCamStr04,SadHigLes07} operate in the regime of large detuning, where MWA can be attributed to collective Rayleigh or Raman scattering. It is clear, however,
that the scattering rate increases as the detuning is decreased \cite{SchTorBoy03,HilKamTar08}, reaching its maximum value at zero detuning. With this in mind, we propose performing MWA  {\it on resonance}, employing a short,
intense pulse for pump 2, thus maximizing the underlying photon emission rate.

Typically, the atoms used  in MWA are cooled well below the critical temperature, $T_c$, for  Bose Einstein condensation, where the spatial coherence length reaches its maximum possible value, i.e. the sample size. For this reason, a BEC is an ideal source of coherent matter waves in much the same way that a laser is an ideal source of coherent light-waves. Despite these considerations, coherent nonlinear mixing of light- and matter-waves does not require a BEC \cite{MooMey01,KetIno01}, and has been demonstrated experimentally for $T>T_c$ \cite{YosTorKug05}. In this work, however, we will focus solely on the BEC regime, with the understanding that many of our conclusions may apply also to thermal gases, as the two systems are indistinguishable for timescales small compared to the coherence lifetime, $\lambda_{dB}/v_r$, where $\lambda_{dB}$ is the thermal de Broglie wavelength, and $v_r$ is the recoil velocity.

While going on-resonance is often associated with rapid heating, and matter-wave decoherence, this is not actually the case in MWA. While going on-resonance certainly increases the decoherence rate, it also increases the MWA gain rate. As both of these rates scale linearly with the background  emission/scattering rate, their ratio, which determines the MWA fidelity, is a universal property, i.e. independent of the detuning and pump strengths for a wide range of parameter settings.
This means that a short, intense control pulse (pump 2), could complete the MWA process on a time-scale much shorter than the spontaneous lifetime, with the net heating effects being no worse than those found in off-resonance MWA.

This approach cannot be applied to Rayleigh scattering MWA schemes, where reabsorption of the idler photons will lead to a cascade of many hundreds of additional matter-wave momentum sidemodes \cite{SchTorBoy03} on the MWA timescale. It therefore applies only to Raman MWA, where a different set of problems arise. In the case of Rayleigh scattering, the AC stark shifts imposed by pump 2 are the same for the signal and pump 1, so that the relative phase between the two matter waves remains constant. In the case of a Raman scheme, however, the AC stark shifts imposed by pump 2 do not match, as different hyperfine states generally have different dipole moments. As a result, the matter-wave relative-phase will oscillate rapidly during the MWA process (typically, the oscillation rate will be $\sim 100$ times faster than the MWA gain rate). This effect, however, does not occur in the resonant case, where the AC Stark shift vanishes.
Thus in addition to providing the largest possible gain, resonant MWA also eliminates the undesirable AC Stark imbalance inherent in Raman MWA

Ultimately, the source of dynamical nonlinearity in MWA is the
atom-field dipole interaction, which becomes a cubic
non-linearity upon second-quantization of the matter-wave field. The interaction between the matter and light fields is primarily
governed by four important parameters: the Rabi frequency,$\Omega$, the detuning $\Delta$,  the spontaneous emission rate, $\Gamma$, and the  optical depth, $D$, which governs the strength of the nonlinearity.

In the far-detuned ($|\Delta|\gg |\Omega|,\Gamma D$), and overdamped ($\Delta=0$, $\Gamma D\gg|\Omega$) regimes, the
excited atomic state is only virtually populated, and can be adiabatically eliminated. This results in
an effective cubic nonlinearity whereby MWA becomes a
four-wave mixing process, unique only  in that the waves are of two different types (2 matter + 2 light).
In the Rabi regime ($\Delta=0$, $|\Omega|\gg\Gamma D$), however, the excited
state is macroscopically populated, so that the system consists of
five dynamically independent waves, which interact via a pair of coupled cubic nonlinearities.
Thus, aside from possible practical applications, resonant MWA may be a system of fundamental interest from a nonlinear-optics perspective.
In fact, treating the excited level dynamically is equivalent to
extending the nonlinear susceptibility to all orders, with the Rabi regime corresponding to the non-perturbative limit of a saturated absorber.
In practice, the resulting strong nonlinearity can lead to nontrivial spatial
effects in both light and matter waves, as  was observed in
a recent Rayleigh superradiance experiment under near-resonant
driving \cite{HilKamTar08}.


\subsection{Preliminary estimates}
\label{prelimest}
Emission of a photon imparts a recoil-kick onto the emitting atom. The solid angle of emission such that the final momentum of the atom lies within the atomic momentum coherence length of a given direction defines a Fresnel mode. Even in  a thermal sample, the atomic momentum coherence-length is non-zero due to the finite size of the sample. For a thermal atomic vapor of width $W$, the corresponding thermal momentum coherence-length is $k_{coh}\sim 1/W$,
so that the Fresnel-mode solid angle is $\Omega_{\cal F}=\pi\left(k_{coh}/k_L\right)^2\sim\pi/(k_LW)^2$. The MWA idler mode corresponds to a Fresnel mode aligned in the direction for which an atom excited from pump 1 is transferred by photon-recoil into the signal mode. In the absence of signal population, the probability of emission into this mode is approximately $\Omega_{\cal F}/4\pi\ll 1$ (neglecting  dipole emission pattern effects). Bosonic stimulation should enhance emission into this mode by the signal-mode population, $(N_s+1)$, relative to all other emission channels, so that the probability for emission into the idler mode is
\beq
P_i=\frac{(N_s+1)\Omega_{\cal F}}{4\pi+N_s\Omega_{\cal F}},
\eeq
with non-idler emission occurring with probability $1-P_i$. The ratio of idler to non-idler emission, which corresponds to MWA gain-to-loss ratio, is thus given by the collective-emission optical depth,
\beq
D=(N_s+1)\frac{\Omega_{\cal F}}{4\pi}\sim \frac{N_s}{A_\perp},
\eeq
where $A_\perp$ is the cross-section of the atomic sample perpendicular to the idler direction. Thus a necessary condition for MWA is a large collective-emission optical depth, $D_s\gg 1$. In terms of the density $n=n_s/(LW^2)$ this gives the standard result $D\sim n\lambda^2L$. Clearly an elongated, ``cigar-shaped'' sample of length $L$ and width $W$, will maximize $D$ for fixed atomic density, provided the signal mode is directed along the long-axis. Thus, at present we consider only condensates with aspect ratios greater than or equal to unity, i.e. $L>W$.

Any Fresnel mode within the geometric solid angle $\Omega_{EF}=\pi(W/L)^2$ of the long-axis is termed an `end-fire mode'. Any mode outside this solid angle has an effective optical depth reduced by the factor $W/L$ relative to that of an end-fire mode. The number of end-fire modes is then given by $M\sim \Omega_{ef}/\Omega_{\cal F}\sim k_L^2W^4/L^2$. This leads us to an important parameter, the `Fresnel number', given by ${\cal F}=k_LW^2/L$, so that $M={\cal F}^2$. A typical cigar-shaped BEC, with an aspect ratio of $10:1$, has ${\cal F}\sim 800$, which does not necessarily correspond to the optimal optical depth for collective emission. For fixed atom number, the optical depth would be increased by increasing the aspect ratio. In doing this, however, one eventually reaches the regime where ${\cal F}\le 1$. At this point, light emitted into an end-fire mode diffracts out of the condensate volume before reaching the far end of the BEC, resulting in a decreased effective optical depth. This effect is characterized by the diffraction length, $L_d=k_Lw^2$, which is the distance a collimated beam of light will propagate before it begins to spread. 

We see that the ratio of the diffraction-length, $L_d$ to the sample length $L$ leads  to the Fresnel number $\cF=k_LW^2/L$. Thus for $1\ll\cF$, the optical depth is sample-length limited, $D\sim n\lambda^2 L$, which can be re-expressed as 
\beq 
\lim_{1\ll\cF}D\sim \frac{N}{\sqrt{v}}\sqrt{\cF},
\eeq
 where $v=k_L^3 LW^2$ is the dimensionless sample volume. For   $\cF\ll 1$, the optical depth is instead diffraction-length limited, $D\sim n\lambda^2 L_d$, which can be re-expressed as
\beq
\lim_{\cF\ll 1}D\sim \frac{N}{\sqrt{v}}\frac{1}{\sqrt{\cF}}.
\eeq
Thus we see that the maximum MWA gain for fixed atom number $N$ is obtained by decreasing the volume while adjusting the geometry such that $\cF\sim 1$, resulting in a maximum optical depth of 
\beq
D_{max}\sim\frac{N}{\sqrt{v}}.
\eeq

The per-atom rate of non-collective photon emission from an atomic sample is given by $P_e  \Gamma$, where $P_e(t)$ is the single-atom excited state probability.
Generalizing this to the nonlinear regime of collective emission into the idler mode, the resulting exponential gain rate for the signal mode occupation is then
\beq G_s=P_e\Gamma D,\label{Gs}\eeq
where $P_e$ is the single-atom excited state probability, and $\Gamma$ is excited-state spontaneous emission rate into the electronic state of the signal matter-wave.
To good approximation, the excited state probability of a driven atom in this system is
\beq
    P_e=\frac{|\Omega|^2}{4\Delta^2+(\Gamma D)^2+2|\Omega|^2},\label{Pesimple}
\eeq
where $\Delta$, and $\Omega$, are the detuning and Rabi frequency, respectively, of pump 2.

By inserting (\ref{Pesimple}) into (\ref{Gs}), we can see that for fixed laser intensity, the MWA gain is always maximized by going on resonance, $\Delta=0$.
Equation (3) shows that the atomic transition saturates in the limit $|\Omega|^2\gg\frac{1}{2}(\Gamma D_s)^2$, leading to the condition
\beq G_s\le\Gamma D_s/2 \label{maxGs},\eeq
as an initial estimate for the upper-limit to the matter-wave amplification rate.
It is important to note, that the saturation intensity has increased by $ D^2_s$, the square of the collective-emission optical depth, due to superradiant broadening of the excited-state linewidth.

The idea to use forward scattering of short, intense, resonant pulses to probe a BEC was first proposed
in 1995 by You, Lewenstein, and Cooper, \cite{YouLweCoo95}, who considered only two-level atoms without recoil in the single spatial-mode approximation.  While this paper established the feasibility of resonantly probing a condensate on a time-scale short compared to the spontaneous lifetime, showing that meaningful phase information could be obtained without decoherence, to the best of our knowledge, no experiments have been performed along these lines. Our proposal for resonant MWA is, from a formal viewpoint, an extension of this idea to the case of a lambda-type three-level system.

\subsection{System-reservoir versus Maxwell-Schr\"odinger models}

There have been two basic models used to describe the MWA
system. The first, pioneered by  Moore and Meystre
\cite{MooMey99b}, is based on extending the Wigner-Weisskopf theory of spontaneous emission to the regime of atomic field theory. In this approach, the light field is eliminated,  resulting in a set of quantum Langevin equations for the atomic field \cite{ZhaWal94}. These equations are
further simplified by quantizing the atomic field onto a quasi-complete set of
`momentum side-modes'
\cite{MusYou00,WanYel05,UysMey08}. We will henceforth refer to this approach as the
quantum coupled mode (QCM) model. In this model, the MWA is attributed to bosonic stimulation by the atoms, whereby recoil into a particular side-mode is enhanced
by a factor of the side-mode occupation number.

The QCM mode, however, fails to explain several
important observations
\cite{PuZhaMey03, SchTorBoy03}. These  include asymmetry between
forward and backward superradiance modes, spatially dependent depletion of the BEC, and subexponential
growth and ringing of the scattered field \cite{ZobNik06,UysMey07}. These effects, usually referred to as
spatial effects or propagation effects, are shown to be important in
the strong driving regime, even though atomic motions can be neglected
\cite{ZobNik05}. In this case, whereas the total atomic density is a constant of motion,  the distribution of probability
over internal energy levels can still depend on position.

In a more recently developed approach aimed at addressing these phenomena, MWA is viewed as
Bragg scattering process with feedback, in which atoms are diffracted by optical
standing waves formed by the pump and idler fields
\cite{TriSha04,ZobNik05,Sha06,ZobNik06,ZobNik07,UysMey07,AveTri08,AveTri08b,HilKamTar08}.
The superradiant enhancement occurs because  the diffraction efficiency
is proportional to the square to grating depth, which depends on the atom
number in the signal wave. This approach incorporates the spatial dependence of the light-matter coupling
via coupled Maxwell-Schr\"odinger equations, which can be efficiently solved at the mean-field level,
hence we term this the semiclassical Maxwell-Schr\"odinger (SMS) model.
As the initial build up of the superradiance from noise is not a mean-field effect, it was treated in the QCM model, which was used to generate
suitable  initial conditions for the  SMS equations. For the case of MWA, however, there is no build-up from noise, so that the dynamics is essentially coherent, thus mean-field theory is expected to work extremely well, as phase fluctuations  are effectively negligible for large enough particle numbers. 

At the mean-field level, and in the single-spatial mode approximation, the QCM equations reduce to three or four dynamical variables, so that the dynamics can be well understood, and accurately described by approximate or exact analytic solutions. Thus we will first explore the QCM results, and then compare and contrast them with the more accurate, but numerically intensive SMS dynamics.

We emphasize that in both models, the light fields adiabatically follow the matter-wave fields, the difference is that only the SMS respects causality,
in the sense that the intensity at position $\vr$ of a field mode propagating in direction $\vk$ can only depend on the integrated atomic polarization in the $-\vk$ direction.
The point is that one should not be misled into thinking that the light fields are independent dynamical variables in the SMS approach.  In general, the coupled atom-field equations  can always be replaced by non-local, but purely atomic non-linear equations using Green's function methods.

\subsection{Organization}
In order to compare the on-resonance dynamics with the previously studied far-detuned dynamics, we will study the MWA
process in the strong-pumping regime where pump laser depletion is a small effect.  We will therefore explore the cross-over between two limiting cases: the Stark regime, $|\Delta|\gg |\Omega|,\Gamma D$, and the Rabi regime, $|\Omega|\gg \Gamma D,\Delta$. In order to illustrate
important underlying physics, we will focus on a simple,
three-level Raman system, with co-propagating pump  and idler light fields.

In section \ref{QFT}, we  use a modified master-equation approach, treating the optical field as a Markovian reservoir, and derive a set of  non-linear non-local equations for the atomic field operators. Then in section \ref{MFT}, we use this approach to  derive the mean-field theory for both atomic and optical fields. Using a Green's function formalism, we show that, at the mean-field level, treating the field as a Markovian reservoir is equivalent to solving coupled Maxwell and Schr\"odinger equations, with the additional benefit of including vacuum-fluctuation effects such as spontaneous emission and the Lamb-shift. We will then compare a theory using the exact Green's function of an ideal dipole with previous approaches to describe MWA and superradiance in the Maxwell-Schr\"odinger picture.

In section \ref{QCM}, we develop the quantum coupled-mode model, based on a single spatial mode approximation for the condensate, and give a precise calculation of the optical depth and dipole-dipole shift . Then, in section \ref{SPE},
we solve the full spatial equations in mean-field theory,  incorporating laser depletion and field-propagation effects.
effects. Lastly, we give a discussion and conclusion in section \ref{ds}.

\section{The Model} \label{model}
A MWA process can be implemented with Raman or Rayleigh transitions,
corresponding to the pump 1 and signal matter waves in different
hyper-fine and/or translational momentum states. In a Rayleigh MWA
process, multiple atomic recoil modes, in addition to the initial pump and signal, would unavoidably appear due to two-photon Bragg scattering of atoms by the pump and idler fields  \cite{SchTorBoy03, ZobNik05,ZobNik06}. In a
Raman MWA, however, such processes are limited by the number of ground-state hyperfine Zeeman sub-levels \cite{WanYel05,
UysMey07}. For this reason, Raman transitions are better suited to MWA applications, such as matter-to-light quantum state transfer.

For concreteness, we consider a very simple Raman MWA configuration, with the pump laser and idler field corresponding to co-propagating right- and left-circularly polarized fields, respectively, i.e. a co-propagating $\Lambda$-system. In this configuration both the pump and signal condensates are at rest ($\bfk\approx 0$). Thus the two condensates must correspond to two different hyperfine $m_F$ states, separated by $\Delta m_F=2$. This co-propagation arrangement is different from the commonly used end-fire configuration where pump 2 and idler beams are
perpendicular \cite{InoPfaGup99,SchTorBoy03,SchCamStr04}. Comparatively, the pump-laser
depletion effect is thus expected to be stronger in our scheme.

As we will see in Sec. \ref{cp}, the maximum obtainable optical depth is given by $D =3.3\times N/\sqrt{k_L^3V}$, where $N$ is the atom number, and $V$ is the volume.  For a typical condensate with $N\sim 10^6$ and $k_L^3V\sim 10^7$, this gives $D\sim 10^3$. The superradiance-broadened lifetime of the excited state is given by $\Gamma D$. For a typical atomic linewidth of $\Gamma\sim 10^7\mbox{Hz}$, this results in  $\Gamma D\sim 10^{10}$, which is larger than the typical excited state hyperfine splitting, $\Delta_{hf}\sim 10^9$. For an alkali metal condensate, which has a single valence electron with a ${^1}S_{1/2}$ ground-state configuration, absorption of a $\sigma^+$ photon and subsequent emission of a $\sigma^-$ photon results in a hyperfine angular momentum change of $\Delta m_F=2$. As flipping the electron spin can only give $\Delta m_F=1$, the co-propagating $\Lambda$-system  requires a change in the nuclear spin of at least $\Delta m_I=1$. As the optical fields do not efficiently couple to spins, this nuclear spin-flip can only be accomplished by the hyperfine interaction when the atom is in the electronically excited ${^1}P$ state. This therefore requires $\Gamma D\ll \Delta_{hf}$, i.e. $D\ll 10^2$. Under such a constraint, incoherent radiation and the corresponding loss of atoms from the pump and signal modes will no longer be negligible. Therefore the standard alkali metal atom condensates are not suitable for resonant MWA experiments.

\begin{figure}
\epsfig{figure=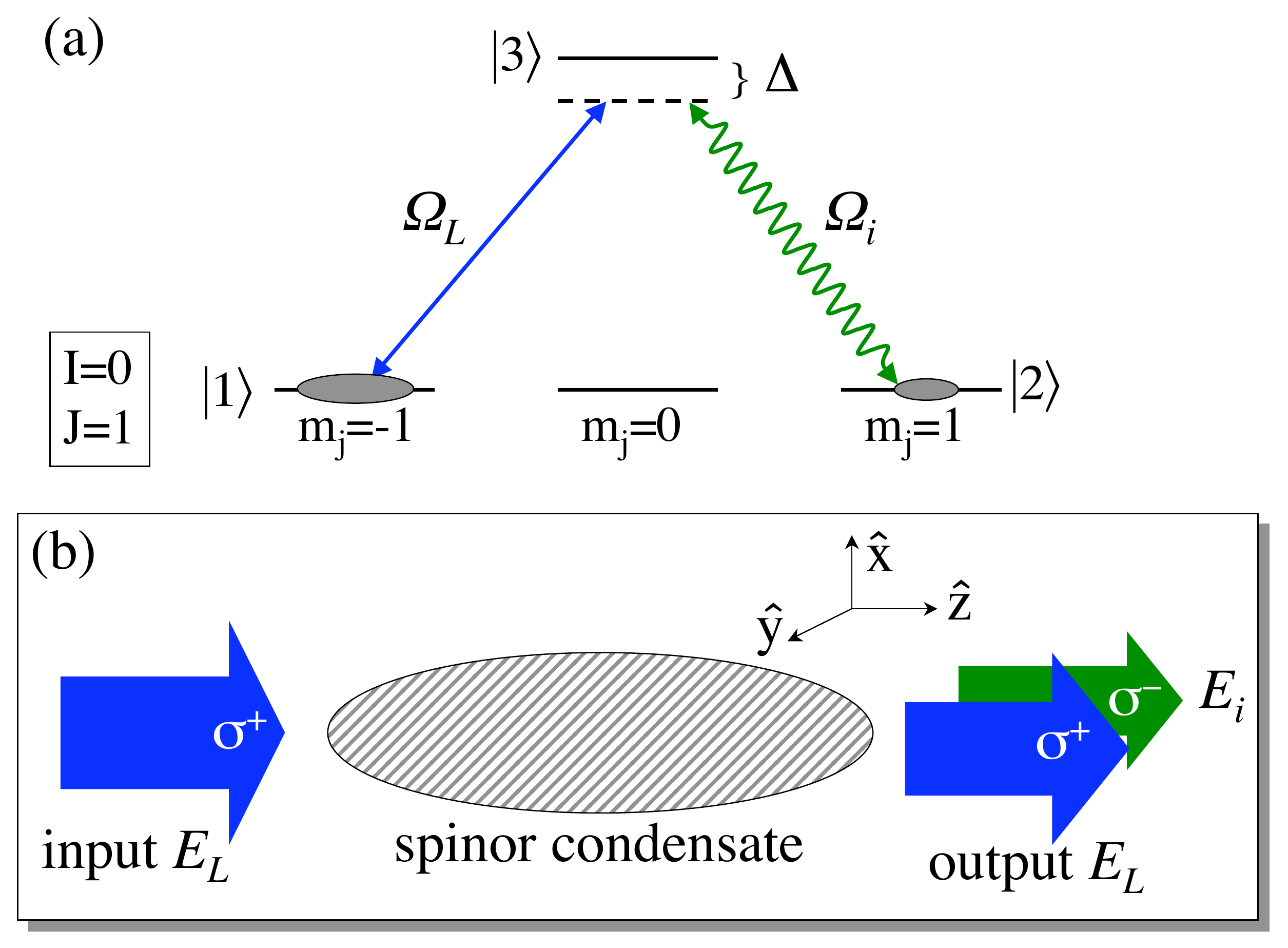, width=8.5 cm} \caption{(Color online) Raman MWA system
configured on a metastable $J{=}1$ state for an isotope with no hyperfine-structure.   Figure (a) shows
the $\Lambda$-level scheme, with the laser and idler fields indicated.  The pump condensate is in internal state $|1\ra$ and the matter-wave signal is in internal state $|2\ra$. The pump laser and idler fields are indicated by their respective Rabi frequencies, $\Omega_L$ and $\Omega_i$. Figure (b) illustrates
the schematic setup for copropagating beams. \label{scheme}}
\end{figure}

The solution  is to choose an atomic species with no nuclear magnetic moment, $I=0$, and therefore no hyperfine structure. To obtain a $\Lambda$-system with co-propagating circularly polarized  fields thus requires a ground state with $J\ge 1$. To avoid sequential re-scattering of the pump and idler fields, an ideal configuration would be $J=1$, with the pump and signal matter-waves corresponding to $m_J=-1$ and $m_J=1$.The level diagram and setup for a MWA scheme based on these considerations is shown in Figures \ref{scheme}a and \ref{scheme}b, respectively.

\begin{figure}
\epsfig{figure=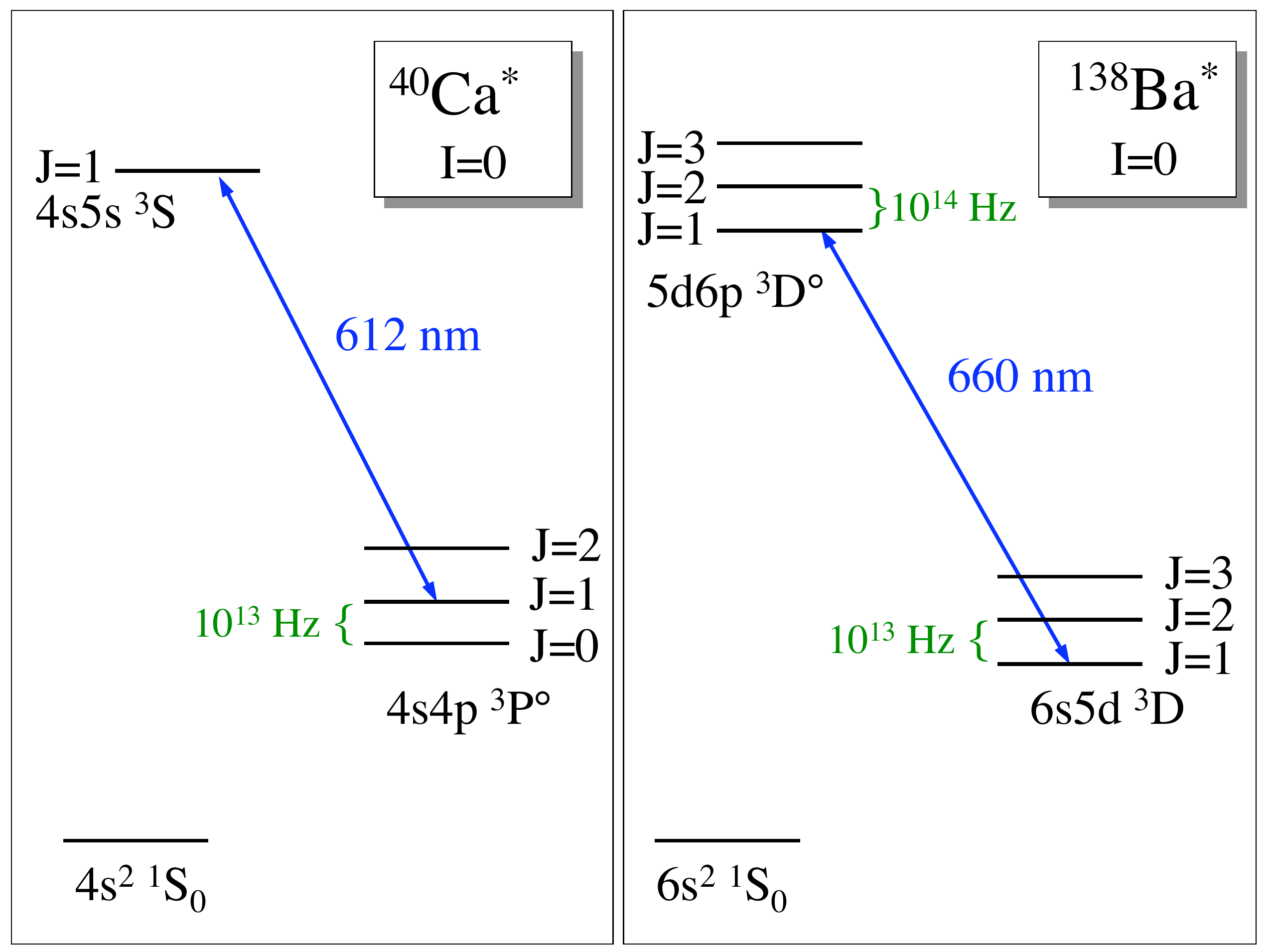, width=8.5 cm} \caption{(Color online) Two viable implementations, one using a metastable triplet P state in $^{40}$Ca and the other using a metastable triplet D state in $^{138}$Ba. By driving a $J=1$ to $J=1$ transition, a pure $\Lambda$ system is realized.
\label{systems}}
\end{figure}

To date, the only atomic species that has been condensed into a $J=1$  state is $^4\mbox{He}^\ast$ \cite{HeBEC1,HeBEC2}, which has a ${^3}\mbox{S}_1$ configuration. Unfortunately, this choice is not viable because the excited-state fine-structure splitting is anomalously small at $\Delta_{fs}\sim 10^{10}\mbox{Hz}$. The best possibility is therefore to prepare a BEC with a species that can be coherently driven into a metastable $J=1$ state. Two atomic species with suitable metastable states for which BEC has been demonstrated are $^{40}\mbox{Ca}$ \cite{CaBEC} and $^{84}\mbox{Sr}$ \cite{SrBEC1,SrBEC2}, both of which have zero nuclear magnetic moment ($I=0$). The 4s4p $^3\mbox{P}^\circ_1$ level of $^{40}\mbox{Ca}$ has a lifetime of $\sim 1\mbox{ms}$, and the 5s5p $^3\mbox{P}^\circ_1$ level of $^{84}\mbox{Sr}$ has a lifetime of $\sim 10\mu\mbox{s}$, both of which would be adequate for resonant MWA, which operates on the time-scale of $\sim1\mbox{ns}$. Another promising species would be $^{138}\mbox{Ba}$, which has been laser cooled and trapped \cite{BaMOT}, with good prospects for achieving BEC \cite{BaBEC}. Decay of its 6s5d level, which has a $^3\mbox{D}_{1,2,3}$ configuration, is doubly-forbidden, and has a measured lifetime of $\sim10\mbox{s}$. For all three of these species, the fine structure splittings are $\sim 10^{13}\mbox{Hz}$, which makes them ideal for resonant MWA experiments. In figure \ref{systems} we show  level structures for potential MWA schemes configured on metastable $J=1$ states in $^{40}\mbox{Ca}$ and $^{138}\mbox{Ba}$. By driving a transition to a $J=1$ excited state, there is no coupling of the pump laser to state $|2\ra$ nor of the idler field to state $|1\ra$, making the system a pure $\Lambda$ configuration.

While a Bose-Einstein condensate (BEC) is not required for coherent collective effects such as MWA or superradiance, we will assume a BEC as a source due to its high optical depth. Later, we will consider the precise temperature requirements imposed by each dynamical regime. In the present scheme, 
both pump and signal modes are at rest, while the excited state propagates at the recoil velocity. 
We presume this will improve the MWA fidelity, which will be important for applications such as state mapping between matter and light waves. We note that in our dynamical simulations, we will neglect the kinetic energy operator for all states, as well as s-wave collisions, based on the assumption that these only become important on time-scales much longer than the fast MWA time-scales we are interested in at present. The effects of s-wave collisions are included only via the initial Thomas-Fermi wavefunction of the BEC.

\subsection{Quantum-field theory for laser driven atoms}
\label{QFT}

The mean-field equations for the atomic field in the QCM model can be derived by treating the electromagnetic vacuum as a Markovian reservoir.  For a generic system, governed by the Hamiltonian, $H_s$, coupled to a zero-temperature bosonic reservoir, governed by 
\beq
H_r=\sum_{\lambda}\int d\bfk \,\omega_\bfk \,a^\dag_{\bfk\lambda}a_{\bfk\lambda},
\eeq
with a coupling described by 
\beq
V=\sum_{\lambda }\int d\bfk\,  (s_{\bfk\lambda }^\dag a_{\bfk\lambda} + a_{\bfk\lambda}^\dag s_{\bfk\lambda }),
\eeq 
where $\{s_{\bfk \lambda}\}$ is any set of system operators. 

We proceed by separating the system operators, $s_{\bfk\lambda}$ into fast and slow parts, relative to the reservoir relaxation time-scale, as
\beq
s_{\bfk\lambda}=\sum_m s_{\bfk\lambda m}e^{-i\omega_mt},
\eeq
where $s_{\bfk\lambda m}$ is slowly-varying, and all fast dynamics is accounted for by the exponential factors. Note that we are {\it not} making the rotating-wave approximation, which would involve keeping only terms for which $\omega_m= 0$.

The equation of motion for the expectation value of an arbitrary system operator, $S$, is then given in the Markoff approximation by
\beqa
&&\!\!\!\!\partial_t \la S\ra =i\la[H_s{+}V_s,S]\ra
+\sum_{\lambda m}\int d\bfk\, \pi\delta(\omega_\bfk-\omega_m)\nn
&&\!\!\times\la 2 s_{\bfk\lambda m}^\dag S s_{\bfk\lambda m}
- s_{\bfk\lambda m}^\dag s_{\bfk\lambda m} S- S s_{\bfk\lambda m}^\dag s_{\bfk\lambda m}\ra,
\label{dSdt}
\eeqa
where 
\beq
V_s=-\sum_{\lambda m}\int d\bfk\, \frac{1}{\omega_\bfk-\omega_m}s_{\bfk\lambda m}^\dag s_{\bfk\lambda m} ,
\eeq
is the system potential induced by the interaction with the reservoir. 

The Hamiltonian for a system of three-level bosonic atoms coupled to the EM-vacuum, and driven on the $|1\ra\to|3\ra$ transition by a laser field, $\bfE_L(\bfr, t)=\bfE_L(\bfr)e^{-i\omega_L t}+\bfE_L\str(\bfr)e^{i\omega_Lt}$, is given in the rotating frame by
\beqa
H_s&=&\Delta\, \int d\bfr\, \Psi_3^\dag(\bfr)\Psi_3(\bfr)\nn
&-&\int d\bfr\, \left[\bfd_1\str\cdot\bfE_L(\bfr)\Psi_1^\dag(\bfr)\Psi_3(\bfr)+H.c.\right],
\eeqa
where $\Delta = \omega_{13}-\omega_L$ is the laser detuning, and $\bfd_j$ is the dipole moment of the $|j\ra\to|3\ra$ transition divided by $\hbar$. The atomic field operators obey the commutation relation $[\Psi_j(\bfr),\Psi_{j'}(\bfr')]=\delta_{jj'}\delta(\bfr-\bfr')$.

The reservoir is characterized in the rotating frame by mode frequencies $\omega_\bfk=c(k-k_L)$, 
 and corresponding system operators 
\beq
s^\dag_{\bfk\lambda 1}=\sum_{j=1,2}\left(\bfd_j\str\cdot\bfe_{\bfk\lambda}\right)\sqrt{\frac{ck}{2\epsilon_0(2\pi)^3}}\int d\bfr\, e^{i\bfk\cdot\bfr}\Psi_3^\dag(\bfr)\Psi_j(\bfr),
\eeq
and
\beq
s^\dag_{\bfk\lambda 2}=\sum_{j=1,2}\left(\bfd_j\cdot\bfe_{\bfk\lambda}\right)
\sqrt{\frac{ck}{2\epsilon_0(2\pi)^3}}
\int d\bfr\, e^{i\bfk\cdot\bfr}\Psi^\dag_j(\bfr)\Psi_3(\bfr) ,
\eeq
with corresponding oscillation frequencies $\omega_1=0$ and $\omega_2=-2c k_L$. The unit-vectors
 $\{\bfe_{\bfk \lambda}\};\ \lambda=1,2$ are any two orthonormal polarization vectors  that are mutually orthogonal  to   $\bfk$.

\subsection{Green's function approach to mean-field theory}
\label{MFT}

Using (\ref{dSdt}) to compute the equation of motion for $\psi_j(\bfr)=\la\Psi_j(\bfr)\ra$, and then making the usual mean-field factorization, we obtain the mean-field theory: 
\beqa
\partial_t\psi_1&=&i\bfd_1\cdot \bfE\str\,\psi_3,\label{dpsi1dt}\\
\partial_t\psi_2&=&i\bfd_2\cdot \bfE\str\,\psi_3,\label{dpsi2dt}\\
\partial_t\psi_3&=&-i\left[\Delta-i\frac{\gamma}{2}\right]\psi_3+i\sum_{j=1,2}\bfd_j\str\cdot\bfE\,\psi_j\label{dpsi3dt},
\eeqa
where $\gamma$ is the excited state spontaneous decay rate, and $\bfE(\bfr)$ is the positive-frequency part of the optical field.  The spontaneous decay term in Eq. (\ref{dpsi3dt}) accounts for the incoherent radiation emitted by the atoms, and results in an irreversible loss of the mean-field density. The optical field, $\bfE$, is then the sum of the laser field and the coherent radiation emitted by the atoms.

The total optical field is the sum of the laser field and a scattered field, $\bfE=\bfE_L+\bfE_s$, where
the scattered field is related to the positive-frequency polarization density,
\beq
\bfP(\bfr)=\sum_j\bfd_j\psi_j\str(\bfr)\psi_3(\bfr),
\eeq
 via a Green's function tensor, $\mathbb{G}(\bfr)$, according to
\beq
\bfE_s(\bfr)=\frac{k_L^2}{\epsilon_0}\int d\bfr'\, \mathbb{G}(\bfr-\bfr')\cdot\bfP(\bfr').
\eeq
Let $\mathbb{G}_0(\bfr)$ be the Green's function tensor inferred  from the equations of motion for the atomic field operators obtained via Eq. (\ref{dSdt}). We find
\beq
\mathbb{G}_0(\bfr)=\frac{k_L}{2(2\pi)^3}\int d\bfk\, (1{-}\hat{\bfk}\,\hat{\bfk})\left[\frac{2k^2}{k^2{-}1}+i\pi\delta(k{-}1)\right]e^{i\bfk\cdot\hat{\bfr}\xi}
,\label{G0intdk}
\eeq
where $\xi=k_L r$.  Here $\hat{\bfk}\,\hat{\bfk}$ is the (outer) tensor-product of the unit-vector $\hat{\bfk}$ with itself.
Performing the $k$-space integration leads to
\beqa
\mathbb{G}_0(\bfr)&=& \frac{e^{i \mathbf{\xi}}}{4\pi r}\left[1{-}\hat{\bfr}\,\hat{\bfr}+(1{-}3\,\hat{\bfr}\,\hat{\bfr})\left(\frac{i}{\xi}-\frac{1}{\xi^2}\right)\right]\nn
&+&\frac{2}{3k_L^2}\delta^3(\bfr).\label{G0}
\eeqa
The field generated by contact term in Eq. (\ref{G0}) corresponds to the internal field of the atom in the ideal-dipole limit \cite{Hannay83}.  To form a proper mean-field theory, we must drop this contact term, as an atom can feel neither its own internal field nor the field inside another atom. Subtracting the internal fields of the atoms is known as the `local-field correction', and has been shown \cite{Bel46,Hannay83} to lead directly to the Clausius-Mossotti/Lorentz-Lorenz relation. The proper Green's function is therefore given by
\beq
\mathbb{G}(\bfr)=\mathbb{G}_0(\bfr)-\frac{2}{3k_L^2}\delta^3(\bfr),
\eeq
which leads to 
\beq
\mathbb{G}(\bfr)=\frac{k_L}{(2\pi)^3}\int d\bfk (1{-}\bfk\,\bfk)\left[\frac{1}{k^2{-}1}+i\pi\delta(k^2{-}1)\right]e^{i\bfk\cdot\hat{\bfr}\xi}.
\label{Gintdk}
\eeq
Performing the $k$-space integration then gives
\beq
\mathbb{G}(\bfr)= \frac{e^{i \xi}}{4\pi r}\left[1-\hat{\bfr}\,\hat{\bfr}+(1-3\,\hat{\bfr}\,\hat{\bfr})\left(\frac{i}{\xi}-\frac{1}{\xi^2}\right)\right],\label{Gr}
\eeq
which reproduces via $\bfE(\bfr)=\mathbb{G}(\bfr)\cdot{\bf p}$, the well-known electric field of an oscillating ideal dipole located at the origin \cite{Jackson}. 

This particular formulation of the Markoff approximation includes retardation effects related to the fast-oscillations of the atomic dipoles at the laser frequency, but neglects retardation related to the slow dynamics of the atomic fields in the rotating frame. This is valid as long as the time-scale of the slow-dynamics is short compared to the retardation time-scale $t_r=L/c$, where $L$ is the system size, and $c$ is the speed of light. For a typical BEC, we have $L\sim 100$ $\mu$m, so that $t_r~1$ ps. In contrast, the fasted time-scale in the rotating frame is given by the super-radiance-enhanced decay rate $\Gamma D$. For typical optical depths of $D\sim 10^3$, this gives a time scale of $\tau \sim 100$ ps. For larger systems, and/or higher optical depths, the theory can be reformulated to take into account all retardation effects.

The mean-field equations of motion, (\ref{dpsi1dt}-\ref{dpsi3dt}), can be expressed as  a set of non-local non-linear equations for the matter waves, given by
\beqa
\partial_t\psi_1&=&\frac{i}{2}\Omega_L\str \psi_3+\frac{i}{2}\sum_{j=1,2} G_{1j}\str[\psi_3\str\psi_j]\psi_3,\\
\partial_t\psi_2&=&\frac{i}{2}\sum_{j=1,2} G_{2j}\str[\psi_3\str\psi_j]\psi_3,\\
\partial_t\psi_3&=&-i\left[\Delta-i\frac{\gamma}{2}\right]\psi_3 \nn
&+&\frac{i}{2}\Omega_L\psi_1+\frac{i}{2}\sum_{j,k=1,2}G_{jk}[\psi_k\str\psi_3]\psi_j,
\eeqa
where $\Omega_L(\bfr)=2\bfd_1\str\cdot\bfE_L(\bfr)$, 
\beq
G_{ij}(\bfr)=\frac{2k_L^2}{\epsilon_0}\, \bfd_i\str\cdot\mathbb{G}(\bfr)\cdot\bfd_j,\label{Gjj}
\eeq 
and we have introduced the compact notation 
\beq
G[f]= \int d\bfr' G(\bfr-\bfr')f(\bfr').
\eeq
 The re-scaled Green's functions, which now have units of $\mbox{s}^{-1}$, are given
For $\bfd_1=\frac{d}{\sqrt{2}}(\hat{\bfx}+i\hat{\bfy})$ and $\bfd_2=\frac{2}{\sqrt{2}}(\hat{\bfy}+i\hat{\bfx})$, by
\beq
G_{jj}(\bfr)=\frac{3\Gamma}{4}\left[1+\frac{z^2}{r^2}+\left(3\frac{z^2}{r^2}-1\right)\left(\frac{i}{\xi}-\frac{1}{\xi^2}\right)\right]\frac{e^{i\xi}}{\xi},
\eeq
\beq
G_{12}(\bfr)=-i\frac{3\Gamma}{4}\frac{\left(x-iy\right)^2}{r^2}\left(1+\frac{3i}{\xi}-\frac{3}{\xi^2}\right)\frac{e^{i\xi}}{\xi},
\eeq
and
\beq
G_{21}(\bfr)=i\frac{3\Gamma}{4}\frac{\left(x+iy\right)^2}{r^2}\left(1+\frac{3i}{\xi}-\frac{3}{\xi^2}\right)\frac{e^{i\xi}}{\xi} ,
\eeq
with
\beq
\Gamma=\frac{d^2k_L^3}{3\pi\epsilon_0}.
\eeq
The diagonal fields, $G_{11}$ and $G_{22}$ describe two-body interactions that conserve atomic hyperfine angular momentum. The off-diagonal fields, $G_{12}$ and $G_{21}$, describe two-body interactions that convert atomic hyperfine angular momentum into center-of-mass orbital angular momentum, with only the total angular momentum being conserved. Only the diagonal fields contain $1/r$ terms, and therefore dominate in extended systems. The initial matter wave signal and pump fields are configured so that the MWA process is driven by these diagonal fields. The off-diagonal fields then describe the generation of vortices during the MWA process. The effects of the off-diagonal fields, however, are expected to be extremely small compared on the MWA time-scale, so that these processes can be ignored.

Treating the off-diagonal fields as a perturbation, and factoring out the rapid spatial oscillations of the excited state and the driving laser via $\psi_3(\bfr)=\psi_e(\bfr)e^{ik_Lz}$,  and $\Omega_L(\bfr,t)=\Omega(\bfr,t) e^{ik_Lz}$, the unperturbed system evolves according to
\beqa
\partial_t\psi_1&=&\frac{i}{2}\left(\Omega\str +G\str[\psi_e\str\psi_1]\right)\psi_e,\label{ptpsi1}\\
\partial_t\psi_2&=&\frac{i}{2}G\str[\psi_e\str\psi_2]\psi_e,\label{ptpsi2}\\
\partial_t\psi_e&=&-i\left[\Delta-i\frac{\gamma}{2}\right]\psi_e+\frac{i}{2}\Omega\psi_1+\frac{i}{2}\sum_{j=1,2}G[\psi_j\str\psi_e]\psi_j,\label{ptpsi3}\nn
\eeqa
where
\beq
G(\bfr)=\frac{3\Gamma e^{ik_L(r-z)}
}{2k_Lr^3}\left[z^2+\frac{r_\perp^2}{2}+\left(z^2-\frac{r_\perp^2}{2}\right)\left(\frac{i}{\xi}-\frac{1}{\xi^2}\right)\right].
\label{G}
\eeq
This set of equations will form the basis for our theoretical description of the MWA process for the present configuration. 

\subsection{Comparison with previous approaches}
Previous theoretical models \cite{ZobNik06,UysMey07} have described spatial propagation effects using coupled Maxwell and Schr\"odinger equations. The advantage of the Markovian approach is that it readily incorporates effects due to electromagnetic vacuum fluctuations, e.g. spontaneous emission. Aside from spontaneous emission, we will now demonstrate that in principle, the two approached give identical results at the mean-field level, the approximation used by previous authors misses fundamentally important physics.

In the Maxwell-Schr\"odinger approach, the matter fields of a two-level atom are governed by 
\beqa
\partial_t\psi_g(\bfr,t)&=&i\bfd\cdot\bfE\str(\bfr,t)\psi_e(\bfr,t),\\
\partial_t\psi_e(\bfr,t)&=&-i\omega_{eg}\psi_e+i\bfd\str\cdot\bfE(\bfr,t)\psi_g(\bfr,t),
\eeqa
while  the electric field is determined by
\beq
\left[\frac{1}{c^2}\partial_t^2-\nabla^2\right]\bfE(\bfr,t)=\bfd \frac{k_L^2}{\epsilon_0}\psi\str_g(\bfr,t)\psi_e(\bfr,t).
\eeq
In order to solve this equation, one can first make the slowly varying envelope approximation,  via
\beqa
\bfE(\bfr, t)&\to&\bfE(\bfr_\perp,z,t) e^{ik_L(z-ct)},\\
\psi_e(\bfr,t)&\to&\psi_e(\bfr_\perp,z,t)e^{ik_L(z-ct)}.
\eeqa
Keeping only the leading order terms gives the standard Paraxial Wave Equation (PWE):
\beq
\left[-2ik_L\left(\frac{1}{c}\partial_t+\partial_z\right)-\nabla_\perp^2\right]\bfE=\bfd \frac{k_L^2}{\epsilon_0}\psi\str_g\psi_e.
\eeq
For small-enough systems, we can neglect retardation, in which case the PWE reduces to
\beq
\left[-2ik_L\partial_z-\nabla_\perp^2\right]\bfE= \bfd \frac{k_L^2}{\epsilon_0}\psi\str_g\psi_e.
\eeq
For a system which is very slowly varying in the perpendicular direction, we can drop the transverse Laplacian, in which case we obtain the Longitudinal Wave Equation (LWE),
\beq
-2ik_L\partial_z\bfE= \bfd \frac{k_L^2}{\epsilon_0}\psi\str_g\psi_e.
\eeq
which yields the equal-time relation
\beqa
\bfE(\bfr_\perp,z)&=&\bfE(\bfr_\perp,z_0)\nn
&+&\!\!i\frac{\bfd k_L}{2\epsilon_0}\int_{z_0}^z dz'\, \psi_g\str(\bfr_\perp,z')\psi_e(\bfr_\perp,z'),
\eeqa
which was used to determine the electric field in previous works \cite{ZobNik06,UysMey07}.

Both the PWE and LWE can be solved using the Green's function formalism. The Paraxial Green's function tensor satisfies
\beq
\left[-2ik_L\partial_z-\nabla_\perp^2\right]\mathbb{G}_P(\bfr)=\delta^3(\bfr),
\eeq
and is given by
\beq
\mathbb{G}_P(\bfr)= \frac{u(z)}{4\pi z}\exp\left[i\frac{k_Lr_\perp^2}{2 z}\right],
\eeq
where $u(z)$ is  the unit-step function. With the definition
$G(\bfr)=\frac{2k_L^2}{\epsilon_0}\bfd\str\cdot\mathbb{G}(\bfr)\cdot\bfd$, this becomes
\beq
\label{GP}
G_P(\bfr)=\frac{3\Gamma}{2}\frac{ u(z)}{k_L z }\exp\left[i\frac{k_Lr_\perp^2}{2z}\right]
\eeq
The longitudinal Green's function must satisfy
\beq
-2ik_L\partial_z\mathbb{G}_L(\bfr)=\delta^3(\bfr),
\eeq
and is given by
\beq
\mathbb{G}_L(\bfr)=i\frac{u(z)}{2k_L}\delta^2(\bfr_\perp),
\eeq
so that
\beq
\label{GL}
G_L(\bfr)=i\frac{3\pi\Gamma}{k_L^2} u(z)\delta^2(\bfr_\perp).
\eeq
We note that the longitudinal Green's function  is purely imaginary. This means that it can describe collective enhancement of spontaneous emission (superradiance) but it does not produce the nonlinear dipole-dipole shift (i.e. the effective two-body interaction potential). The paraxial Green's function, on the other hand, has both real and imaginary parts, and therefore captures the essential physics of the light-matter interaction, differing from the full Green's function only in its degree of precision.

The Paraxial Green's function can be derived as an approximation to the full Green's function (\ref{G}) by expanding separately the pre-factor and exponent to leading order in powers of $r_\perp/z$. which gives
\beq
G(\bfr)\approx\frac{3\Gamma}{2 k_L|z|}\exp\left[i\frac{ k_L r_\perp^2}{2|z|}\right] e^{ik_L(|z|-z)}.
\eeq
Note that the phase factor $e^{ik_L(|z|-z)}$ becomes unity for $z>0$ and $e^{-i2k_Lz}$ for $z<0$. Thus, under integration we have 
\beq
e^{ik_L(|z|-z)}\approx u(z), 
\eeq
so that we obtain the paraxial Green's function of Eq. (\ref{GP}) as the leading-order expansion of $G(\bfr)$ is powers of $r_\perp/z$. Thus we should expect the paraxial Green's function to give good accuracy for condensates with large aspect ratios, and hence small ${\cal F}$, with the accuracy decreasing as ${\cal F}$ increases. 

To obtain the Longitudinal Green's function, we assume that the polarization density is approximately uniform on the scale of the optical wavelength, and replace the  $G_P(\bfr)$ and replace the transverse exponential with a transverse delta function, 
\beq
\exp\left[i\frac{k_Lr_\perp^2}{2z}\right]\approx g \delta^2(\bfr_\perp),
\eeq
where
\beq
g=\int d\bfr_\perp \exp\left[i\frac{k_Lr_\perp^2}{2z}\right]= i\frac{2\pi z}{k_L},
\eeq
which leads directly to Eq. (\ref{GL}).

In section \ref{SPE}, we will compare the mean-field results for the full Green's function to those given by the Paraxial and Longitudinal approximate Green's functions and the single spatial-mode approximation presented in section \ref{QCM}.

\section{Quantum Coupled-Mode theory }
\label{QCM}

In this section, we develop the quantum coupled-mode (QCM)
theory for the three-level Raman MWA system. In section \ref{modes}, we take the mean-field equations (\ref{ptpsi1}-\ref{ptpsi3}) and expand them onto a truncated set of spatial modes defined by the Thomas-Fermi (TF) density profile of a condensate. In section \ref{cp}, we then evaluate the complex collectivity-parameter which governs the MWA process. Because the TF density is a very accurate description, we expect quantitative agreement between the calculated parameter and experiment. In section
\ref{QCManalytics}, we develop approximate analytic solutions to the QCM equations of motion, for each of the three main dynamical regimes, which are then compared with exact numerical solutions of the QCM equatons in section \ref{nr}.\\

\subsection{Spatial mode Expansion}
\label{modes}

In practice, the mean-field equations (\ref{ptpsi1}-\ref{ptpsi3})  are difficult to solve numerically, due to their non-local nature, and the singularities in $G(\bfr)$. 
For short times, it may be valid to neglect the spatial dependences of the mean-field parameters and invoke the single-mode approximation.
This is accomplished by introducing the mode amplitudes, $c_1$, $c_s$, and $c_e$, via
\begin{eqnarray}
    \psi_1(\bfr,t) &=& \sqrt{N} c_1(t)\phi(\bfr) \\
    \psi_2(\bfr,t)&=&\sqrt{N} c_s(t)\phi(\bfr)\\
   \psi_3(\bfr,t) &=& \sqrt{N} c_e(t) \phi(\bfr)e^{i\bfk_L\cdot\bfr},
\end{eqnarray}
where $N$ is the initial atom number, $\phi(\bfr)$ is the initial wavefunction of the condensate, and
$\bfk_L=k_L\hat{\bfz}$ the wavevector of pump 2.  The atomic population in
the pump mode is then $N_1(t)=N|c_1(t)|^2$, and the signal mode population is $N_s(t)=N|c_s(t)|^2$. 

After inserting the mode expansion, the mean-field theory reduces to
\beqa
\label{mf1}
   \ppt c_1 &=&  \frac{i}{2}\Omega_L\str c_e+\frac{1}{2}\Gamma f\str N |c_e|^2 c_1 \\
\label{mf2}
     \ppt c_s &=&\frac{1}{2}\Gamma f\str N|c_e|^2 c_s \\
\label{mfe}
     \ppt c_e &=& -i(\Delta-i\gamma/2) c_e+\frac{i}{2}\Omega_L c_1-\frac{\Gamma}{2} f N\sum_{j=1,s}|c_j|^2 c_e,\nn
\eeqa
where the collectivity parameter, $f$, is given by
\begin{equation}
\label{f}
f=\frac{1}{i\Gamma}\int d^3r d^3r'
|\phi(\bfr)|^2|\phi(\bfr')|^2 G(\bfr-\bfr')
e^{-ik_L(z-z')}.
\end{equation}
We note that have has both a real and an imaginary part. The real part,  $f_R$ leads to enhanced decay of the excited state, with commensurate feeding of the two ground states. The optical depth is therefore given by $D=f_RN$. The imaginary part, $f_I$ leads to cross-phase modulation between the excited and ground states, thus imprinting a nonlinear phase-shift onto the signal mode. Introducing $\chi$ as the ratio between the amplitudes of the real and imaginary parts, we have,
\beq
f=f_R(1-i\chi).
\eeq 
With the dimensionless variables $\tau=\Gamma D t$, $\tilde\Omega=\Omega/\Gamma D$, $\tilde\gamma=\gamma/\Gamma D$, and $\tilde\Delta=\Delta/\Gamma D$, the equations of motion become
\beqa
\pptau c_1&=&\frac{i}{2}\tilde\Omega\str c_e+\frac{1}{2}\left(1+i\chi\right)|c_e|^2c_1\label{dc1dtau}\\
\pptau c_s&=&\frac{1}{2}\left(1+i\chi\right)|c_e|^2c_s\label{dcsdtau}\\
\pptau c_e&=&-i(\tilde\Delta-i\tilde\gamma/2)c_e+\frac{i}{2}\tilde\Omega c_1\nn
&&-\frac{1}{2}(1-i\chi)\sum_{j=1,s}|c_j|^2c_e.\label{dcedtau}
\eeqa
These equations form the basis of the quantum coupled-mode (QCM) model.

\subsection{Collectivity parameter}
\label{cp}
Within the QCM framework, the strength of the nonlinearity is governed by the collectivity parameter, $f$, which depends only on the condensate geometry.
By combining Eqs. (\ref{Gintdk}) and (\ref{f}), we can express the collectivity parameter as
\beq
\label{fS}
f=\frac{3}{8\pi^2}\int d\bfk\, (2{-}k_\perp^2)\left[\pi\delta(k^2{-}1){-}\frac{i}{k^2{-}1}\right]S\left(k_L(\bfk{-}\hat{\bfz})\right),
\eeq
where
\beq
\label{Sk}
S(\bfk)=\left|\int d^3r |\phi(\bfr)|^2 e^{-i\bfk\cdot\bfr}\right|^2
\eeq
is the static structure function of the condensate. 

The spatial profile of the
condensate is generally given to excellent approximation by the Thomas-Fermi distribution,
\begin{equation}
\label{TF}
  |\phi(\bfr)|^2=\frac{15}{\pi W^2L}\left[1-\left(\frac{2\rho}{W}\right)^2-\left(\frac{2z}{L}\right)^2\right],
\end{equation}
where $W$ and $L$ are width and length of the condensate. From Eqs. (\ref{TF}) and (\ref{Sk}), it is straightforward to show that for the Thomas-Fermi density profile, the static structure function is given by
$S(\bfk)=\left|p\left(K(\bfk)\right)\right|^2$, where
\beq
p(K)=\left(\frac{15}{K^3}-\frac{45}{K^5}\right)\sin\left(K\right)
+\frac{45}{K^4}\cos\left(K\right),
\eeq
and
\beq
K=\sqrt{\left(\frac{k_xW}{2}\right)^2+\left(\frac{k_yW}{2}\right)^2+\left(\frac{k_zL}{2}\right)^2}.
\eeq
It is logical to parametrize the condensate dimensions, $L$ and $W$, by the Fresnel number, $\cF=k_LW^2/L$, and the dimensionless volume $v=k_L^3LW^2$, so that we can consider changing the geometry of the TF density profile while holding the mean-density fixed. With $(k_LW)^2=\sqrt{v\cF}$ and $k_L L=\sqrt{v/\cF}$,  and
with the change of variables from $\{k_z,k_\perp\}\to\{x,u\}$, related by $k_\perp=\frac{2}{\sqrt{v\cF}}x \sqrt{1{-}u^2}$, and $k_z{-}1=2\sqrt{\frac{\cF}{v}}xu$, this becomes
\beq
f=\frac{3}{\pi v}
\int_0^\infty dx\, \left|p(x)\right|^2\big(g(v,\cF,x)-i h(v,\cF,x)\big),
\eeq
where
\beq
g(v,\cF,x)=\frac{\pi}{y}\frac{\sqrt{v\cF}-2x^2(1-u_-^2)}{(u_+-u_-)},
\eeq
and
\beqa
h(v,\cF,x)&=&-\frac{4x^2}{1-\frac{\cF^3}{v}}-\frac{\left(\sqrt{v\cF}-2x^2(1-u_-^2)\right)}{(u_+-u_-)\left(1-\frac{\cF^3}{v}\right)}\nn
&&\times\ln\left[\frac{u_+-u_-+u_+u_--1}{u_+-u_--u_+u_-+1}\right]
\eeqa
with $y=1{-}\cF\sqrt{\frac{\cF}{v}}$, and $u_\pm=\frac{\cF\pm\sqrt{\cF^2+4yx^2}}{2yx}$.

\subsubsection{Real part: collective enhancement factor}
The real part of the collectivity parameter governs the short-time MWA rate, and is given by
\beq
f_R=\frac{3}{\pi v}\int_0^\infty dx\,  \left|p(x)\right|^2\, g(v,\cF,x).
\label{fR}
\eeq
Expanding to leading order in $1/v$ gives the large-volume limit
\beq
\lim_{v\to\infty}f_R=3\sqrt{\frac{\cF}{v}}\int_0^\infty dx\,\frac{ x  \left|p(x)\right|^2}{\sqrt{\cF^2+4x^2}}.
\label{fRbigv}
\eeq
We see that in this limit, the collective gain times the square-root of the volume, $\sqrt{v}f_R$, is a universal parameter, depending only on the Fresnel number, ${\cal F}$. 
Expanding to leading order in $1/\cF$ gives the $\cF\gg 1$ limiting behavior,
\beq
\lim_{{\scriptstyle v\to\infty}\atop{\scriptstyle \cF\to\infty}}f_R=\frac{3}{\sqrt{v\cF}}\int_0^\infty dx\, x \left|p(x)\right|^2=\frac{75}{8\sqrt{v\cF}},
\eeq 
which gives the standard scaling of $D\sim n \lambda^2L$ for the optical depth, $D=f_RN$.
In the opposite limit of a very small Fresnel number, $\cF\ll 1$, which corresponds necessarily to a large aspect ratio,  $f_R$ has the limiting form
\beq
\lim_{{\scriptstyle v\to\infty}\atop{\scriptstyle \cF\to 0}}f_R=\frac{3}{2}\sqrt{\frac{\cF}{v}}\int_0^\infty dx\,\left|p(x)\right|^2
=\frac{15\pi}{14\sqrt{v\cF}}\cF,
\eeq 
which gives the expected scaling of $D\sim n\lambda^2 L_d$, where $L_d=W^2/\lambda$ is the diffraction length.

\begin{figure}
\epsfig{figure=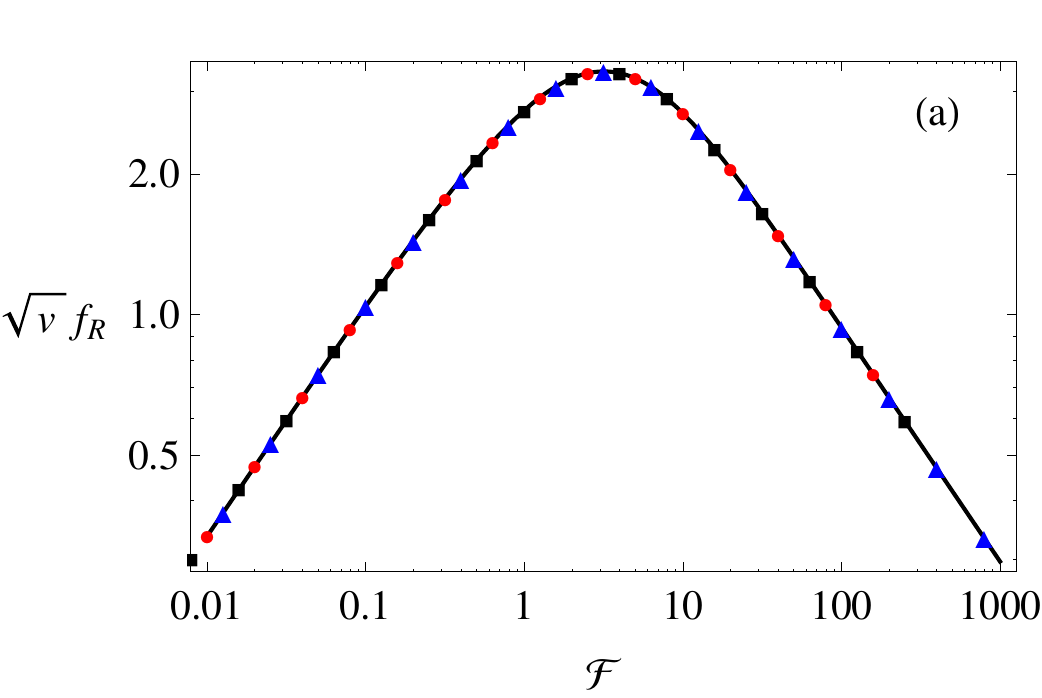,width=8.5cm}
\epsfig{figure=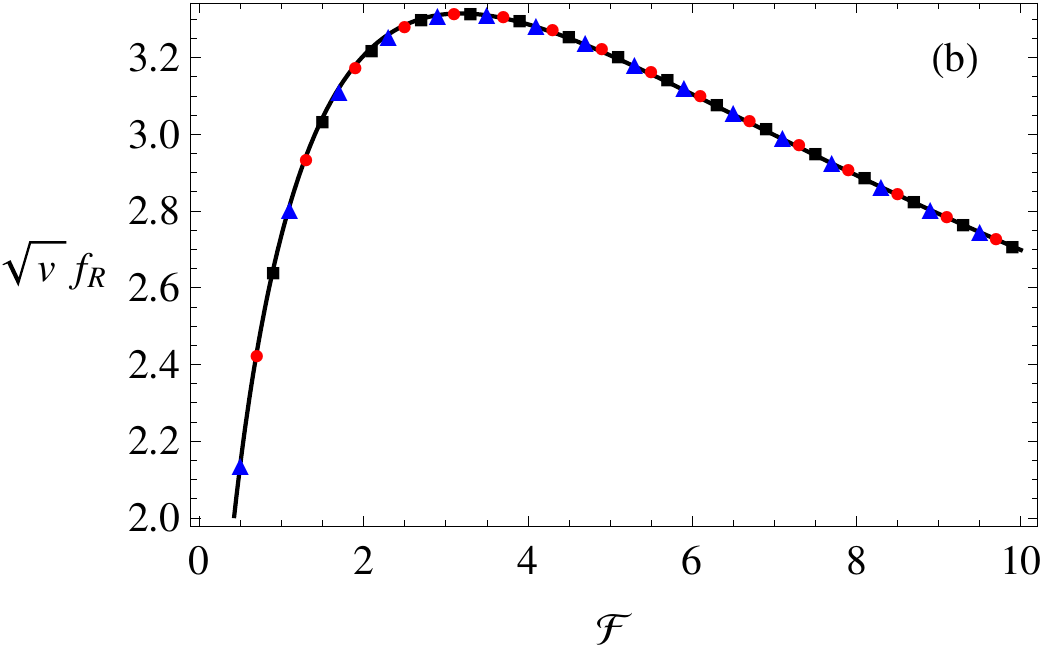,width=8.5cm}
\caption{(Color online) The real part of cooperative parameter, $f_R$, as a function of the Fresnel number, $\cF$. Discrete points are numerical results for the Thomas-Fermi profile (\ref{TF}) for finite volumes, while the solid line corresponds to numerical results for $v=\infty$. The (red)circles, (black) squares, and (blue) triangles correspond to $v=10^7$, $v=10^8$, and $v=10^9$, respectively. \label{fRF}}
\end{figure}

The two limits
can  be merged via the Pad\'e  approximant
\beq
f_R\approx c_0\sqrt{\frac{\cF}{v}}\frac{1+c_1\cF}{1+c_2\cF+c_3\cF^2},
\label{fRPade}
\eeq
with $c_0=3.3660$, $c_1=0.1256$,  $c_2=0.3390$, and $c_3=4.510\times 10^{-2}$, determined from the asymptotic formulas and numerical fitting in the cross-over regime, which gives excellent agreement with the exact expression (\ref{fR}). This is illustrated in Fig. \ref{fRF}, where we plot $\sqrt{v}f_R$ versus $F$ for $v=\infty$.  Figure \ref{fRF}a shows a wide range of $\cF$-values on a log-log scale, whereas Fig. \ref{fRF}b shows the crossover region on a linear scale.  We note that $\sqrt{v}f_R$ is effectively independent of $v$ for $v\gg 10^5$, which is generally satisfied for typical condensates containing $N\sim 10^6$ atoms.

We see that for fixed condensate volume, the coefficient $f_R$, and therefore the optical depth for collective emission, $D=f_RN$, is maximized at $\cF=3.181$, corresponding to an aspect ratio of $\cA=0.420 v^{1/4}$. For $v=10^7$, this gives $\cA=23.6$ as the optimal aspect ratio with respect to the collectivity of the MWA process. For $N=10^6$, the corresponding maximum optical depth is then $D_{max}=1050$. For an aspect ratio of $\cA=10$, at $v=10^7$, we have instead $\cF=10$, and $D=853$. For spherical geometry, we have instead $\cF=215$ and $D=202$.

\subsubsection{Imaginary part: cross-phase modulation}

The imaginary part of the collectivity parameter is proportional to the induced dipole-dipole interaction  energy of the sample. From a nonlinear optics perspective, we
see from Eqs. (\ref{mf1}), (\ref{mf2}), and (\ref{mfe}), this energy manifests itself as a source of cross-phase modulation between the ground and excited states. The magnitude of the imaginary part of $f$ is given by
\beq
\label{fITF}
 f_{I}=\frac{3}{\pi v}\int_0^\infty dx\, \left|P(x)\right|^2 h(v,\cF,x).
\eeq
Expanding the integrand to leading order in powers of $1/v$ gives,
\beq
\lim_{v\to\infty}f_I=\frac{6}{\pi}\sqrt{\frac{\cF}{v}}\int_0^\infty dx\,x\left|p(x)\right|^2 \frac{\ln\left[\frac{\cF}{\sqrt{\cF^2+4x^2}-2x}\right]}{\sqrt{\cF^2+4x^2}}.
\eeq
Again, we see that $\sqrt{v}f_I$ is a universal parameter, depending only on ${\cal F}$ in the large-volume limit.
Taking the limit $\cF\to\infty$ gives
\beq
\lim_{{\scriptstyle v\to\infty}\atop{\scriptstyle \cF\to\infty}}f_I=\frac{12}{\pi\sqrt{v\cF}\cF}\int_0^\infty dx\, x^2 \left|p(x)\right|^2
=\frac{180}{7\sqrt{v\cF}}\frac{1}{\cF},
\eeq
scaling therefore as $L/W^4$. Similarly, taking the limit $\cF\to 0$, corresponding to a very large aspect ratio, gives
\beqa
\lim_{{\scriptstyle v\to\infty}\atop{\scriptstyle \cF\to0}}f_I&=&\frac{3}{\pi}\sqrt{\frac{\cF}{v}}\int_0^\infty dx\, \left|p(x)\right|^2\ln\frac{4x}{\cF}\nn
&=&\frac{15\cF}{7\sqrt{v\cF}}\left(\frac{1517}{1260}-\gamma_e+\ln2-\ln\cF\right),
\eeqa
whose pre-factor scales as $1/L$, but which diverges logarithmically with decreasing $\cF$.

\begin{figure}
\epsfig{figure=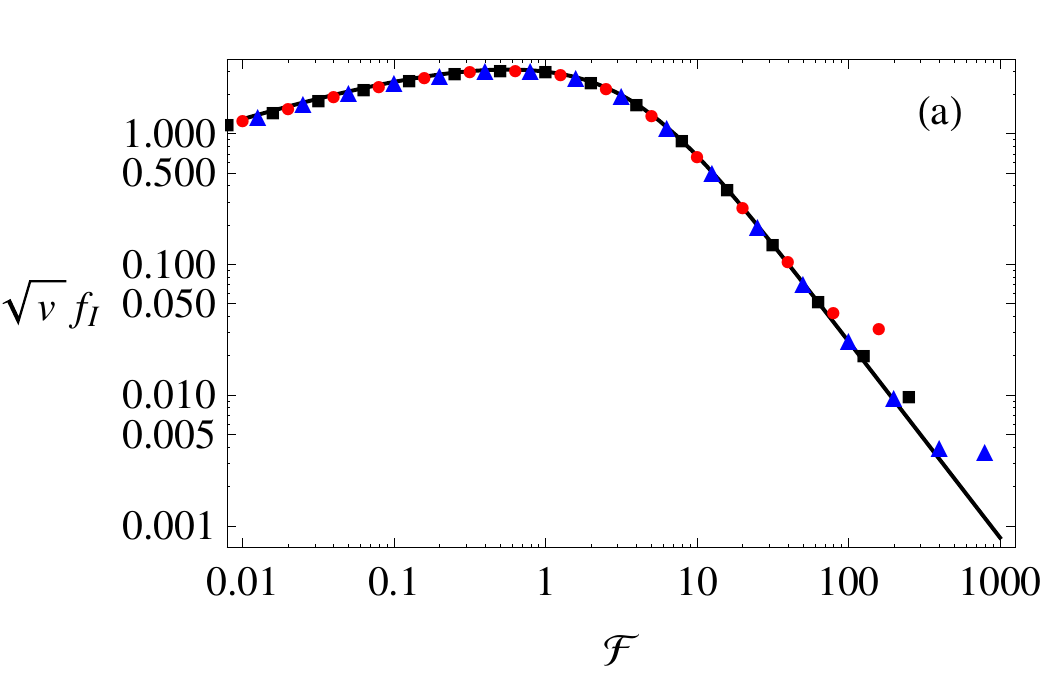,width=8.5cm}
\epsfig{figure=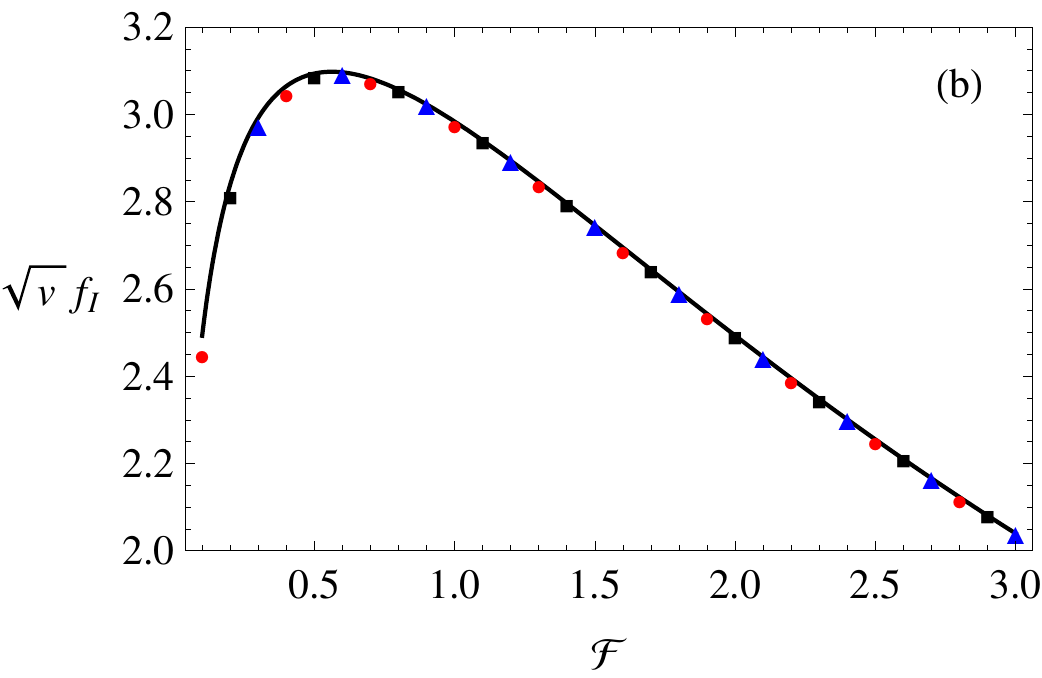,width=8.5cm}
\caption{(Color online) The imaginary part of cooperative parameter, $f_I$, as a function of the Fresnel number, $\cF$. The discrete points are numerical results for the Thomas-Fermi profile (\ref{TF}) for finite volumes, while the solid line corresponds to the numerical result for $v=\infty$. The (red)circles, (black) squares, and (blue) triangles correspond to $v=10^7$, $v=10^8$, and $v=10^9$, respectively. \label{fig3}}
\end{figure}

The two limits can be joined by the Pad\'e approximant,
\beqa
f_I\approx\frac{15\sqrt{\cF}}{7\sqrt{v}}\frac{\left(d_0{-}\ln\frac{\cF}{1+\cF}\right)(1+d_1\cF)}
{1+d_2\cF+d_3\cF^2+d_4\cF^3},
\label{fIPade}
\eeqa
where $d_0=1.453$, $d_1=0.09402$, $d_2=1.223$, $d_3=0.5531$, $d_4=0.1215$, and $d_5=0.01138$.
This is validated in Fig. \ref{fig3} where we plot $\sqrt{v}f_I$ versus $\cF$ for $v=10^7,10^8,10^9$.
Figure \ref{fig3}a shows a wide range of $\cF$-values on a log-log scale, while Fig. \ref{fig3}b shows the cross-over region on a linear scale.

\begin{figure}
\epsfig{figure=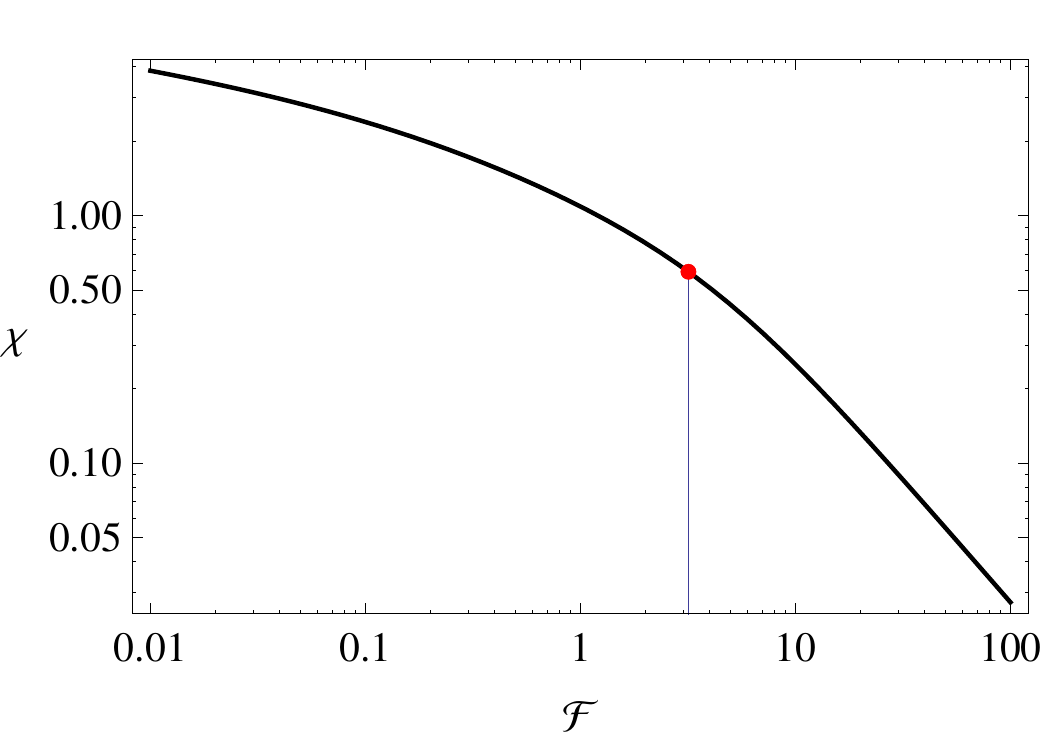,width=8.0cm}
\caption{(Color online) The nonlinear coefficient $\chi=f_I/f_R$ versus Fresnel number, $\cF$ on a log-log scale. The solid line gives the numerical result for $v=\infty$, and the red dot indicates the location of the maximum $f_R$ at $\cF=3.18$. \label{chifig}}
\end{figure}

The strength of cross-phase modulation relative to the collective gain is given by $\chi=f_I/f_R$. In the limit $\cF\ll 1$, corresponding to a highly-elongated condensate, the asymptotic form of this parameter is
\beq
\chi\approx 2\pi(1.453-\ln\cF)\gg 1.
\eeq
In the opposite regime, $\cF\gg 1$, we have instead
\beq
\chi\approx \frac{24}{\pi\cF},
\eeq
which is small provided $\cF\gg 8$. The full behavior of $\chi$ is shown in Fig. \ref{chifig}, where we plot the ratio of the Pad\'e approximants (\ref{fIPade}) and (\ref{fRPade}) versus $\cF$.  We see that $\chi$ decreases monotonically with increasing $\cF$, which corresponds to a {\it decreasing} aspect ratio.

\subsection{QCM dynamics: analytic approximations}
\label{QCManalytics}
In this section, we study the dynamics of the Raman MWA
system by solving the QCM equations (\ref{mf1})-(\ref{mfe}), neglecting atom losses from the pump and signal mode space due to spontaneous emission. In the absence of losses,  the QCM equations become
\beqa
\ppt c_1&=&i\frac{\tilde\Omega\str}{2}c_e+\frac{1}{2}\left(1+i\chi\right)|c_e|^2c_1\label{dc1dtaunoloss}\\
\ppt c_s&=&\frac{1}{2}\left(1+i\chi\right)|c_e|^2c_s\label{dcsdtaunoloss}\\
\ppt c_e&=&\!\!\!-i\tilde\Delta\,c_e+i\frac{\tilde\Omega}{2}c_1
-\frac{1}{2}\left(1-i\chi\right)\sum_{j=1,2}|c_j|^2c_e.\label{dcedtaunoloss}
\eeqa
We will now derive approximate analytic solutions for both the far-detuned Stark and near-resonance Rabi-regime.

\subsubsection{Stark regime}

In the Stark regime, defined as $|\Omega|^2+\Gamma^2D^2\ll \Delta^2$, the excited state amplitude adiabatically follows the pump-matter-wave amplitude. For negligible losses, we than have $|c_2|^2\approx 0$ and $|c_1|^2+|c_s|^2\approx 1$. The adiabatic solution for $c_e$ is then
\beq
c_e\approx \frac{\tilde\Omega c_1}{2\tilde\Delta-\chi-i},
\eeq
which leads to
\beqa
\ddtau c_1&=&\tilde{R}\left[i\tilde\Delta-\frac{1}{2}(1+i\chi)|c_s|^2\right]\,c_1,\label{dc1dtauStark}\\
\ddtau c_s&=&\frac{\tilde{R}}{2}\left(1+i\chi\right)|c_1|^2c_s,\label{dcsdtauStark}
\eeqa
where 
\beq
\tilde{R}=\frac{|\tilde\Omega^2|}{(2\tilde\Delta-\chi)^2+1},
\label{Rweak1}
\eeq
 is the dimensionless MWA rate constant.
Using $|c_1|^2\approx 1-|c_s|^2$, this leads to the rate equation
\beq
\ddtau n_s=\tilde{R}(1-n_s)n_s,
\eeq
where $n_s=|c_s|^2$ is the  fractional population of the signal mode.

This has the analytic solution,
\beq
n_s(t)=\frac{n_s(0)  e^{R t}}{1+n_s(0)\left(e^{R t}-1\right)},
\label{nst}
\eeq
where $R$ is the MWA rate, given in original units by
\beq
R\approx \frac{|\Omega^2|\Gamma D}{(2\Delta-\chi\Gamma D)^2+\Gamma^2D^2}.
\eeq
This shows that the  true resonance condition  is $\Delta=\chi\Gamma D/2$, corresponding to $\omega_L=\omega_e-\omega_1-\frac{\chi}{2}\Gamma D$. This represents a density-dependent red-shift in the resonance due to the atomic dipole-dipole interaction. The FWHM of the resonance is $\Gamma D$, which is larger than the natural linewidth, $\Gamma$, due to superradiance-broadening.

Expanding to leading-order in the detuning, we obtain the Stark-regime MWA rate,
\beq
R_S=\frac{|\Omega|^2\Gamma D}{4\Delta^2},
\eeq
which is larger than the  spontaneous photon-scattering rate by a factor of the optical depth, $D$, characteristic of a collective emission process.

From Eq. (\ref{nst}) we see that for small initial signal fraction, $n_s(0)\ll 1$, there are two distinct time-scales. For times short-enough such that $n_s(0)e^{Rt}\ll1$, depletion of the pump condensate is negligible. In this regime, we have
\beq
n_s(t)\approx n_s(0)e^{Rt},
\eeq
so that $R$ is clearly identified as the MWA exponential gain-rate. For longer times, such that $n_s(0)e^{Rt}\gg 1$, the pump condensate is depleted, resulting in the saturated response $\lim_{t\to\infty}n_s(t)=1$.

The signal-mode phase-shift, defined via $c_s=\sqrt{n_s}e^{i\phi_s (t)}$, is governed by
\beq
\ddtau \phi_s=\frac{\chi}{2}\tilde{R}(1-n_s),
\eeq
which yields the analytic solution
\beq
\phi_s(t)=\frac{\chi}{2}\left[Rt-\ln\left(n_1(0)+n_s(0)e^{Rt}\right)\right].
\label{phist}
\eeq
This represents a nonlinear phase-shift that can be attributed to the inter-atomic dipole-dipole interaction. In the limit $t\to\infty$, we find the net MWA phase shift is given by
\beq
\lim_{t\to\infty}\phi_s(t)=\frac{\chi}{2}\ln \frac{1}{n_s(0)}.
\label{phisinfty}
\eeq
The sign of the phase-shift is positive, with a magnitude proportional to the log of the initial signal-mode population fraction, hence a smaller seed-value results in a larger phase-shift. This is readily interpreted as being due to a longer amplification time during which to accumulate the phase shift. In the weak pumping regime, $|\tilde\Omega|^2\ll (2\tilde\Delta+\chi)^2+1$, the limiting value of the MWA phase-shift is independent of the detuning and pump intensity, depending only on the dimensionless parameter $\chi$.

\subsubsection{Rabi Regime}

For the case $\tilde\Omega^2\gg (\tilde\Delta-\chi/2)^2$, the atomic transition will be saturated, and the the atoms will exhibit Rabi-oscillations between states $|1\ra$ and $|e\ra$. To develop  analytic approximations for this regime, we first introduce the dressed-state amplitudes
\beq
c_\pm=\frac{1}{\sqrt{2}}(c_1\pm c_e)e^{\mp i\tilde\Omega\tau/2},\label{cpm}
\eeq
which satisfy the initial conditions $c_\pm(0)=\sqrt{n_1(0)/2}$. We can construct secular equations of motion for $c_\pm(t)$ by dropping all fast-rotating terms, yielding
\beqa
\ddtau c_\pm&=&-\frac{i}{2}\tilde\Delta c_\pm-\frac{1}{4}|c_s|^2c_\pm\nn
&+&i\frac{\chi}{4}(|c_s|^2+|c_\pm|^2)c_\pm,\label{dcpmdtauRabi}\\
\ddtau c_s&=&\frac{1}{4}(1+i\chi)(|c_+|^2+|c_-|^2)c_s.\label{dcsdtauRabi}
\eeqa
Taking into account the conservation law $|c_+|^2+|c_-|^2+|c_s|^2=1$, this leads directly to
\beq
\ddtau n_s= \frac{1}{2}(1-n_s)n_s,
\eeq
and
\beq
\ddtau\phi_s=\frac{\chi}{4}(1-n_s),
\eeq
Which reproduce the results of Eqs. (\ref{nst}), (\ref{phist}), and (\ref{phisinfty}), but with the Rabi-regime MWA rate being given instead by
\beq
R_R=\frac{\Gamma D}{2}.
\eeq
This shows that the signal-mode amplitude and phase dynamics in the QCM model are universal as the detuning is swept across resonance, governed by a single rate constant,
\beq
R=\frac{|\Omega|^2\Gamma D}{(2\Delta-\chi\Gamma D)^2+\Gamma^2 D^2+2|\Omega^2|}.
\label{R}
\eeq

\subsubsection{Numerics and Comparisons}
\label{nr}

To verify the preceding analysis, we numerically solve the
mean-field equations (\ref{dc1dtau})-(\ref{dcsdtau}), for the case $N=10^6$ and $v=10^7$. For the Stark regime, we take $\tilde\Delta=10$, and $\tilde\Omega=1$, and choose ${\cal F}=10$, corresponding to an aspect ratio of $A=10$ and an optical depth of $D=853$. For the Rabi regime, we take $\tilde\Delta=0$, $\tilde\Omega=10$, and choose ${\cal F}=3.18$, corresponding to an aspect ratio of $A=24$ and a maximized optical depth of $D=1048$. We consider two sets of initial conditions, $n_1(0)=0.9$, $n_s(0)=0.1$, $n_e(0)=0$, corresponding to a typical amplification experiment; and $n_1(0)=n_s(0)=0.5$, $n_e(0)=0$, corresponding to a state-transfer experiment.

\begin{figure}
\epsfig{figure=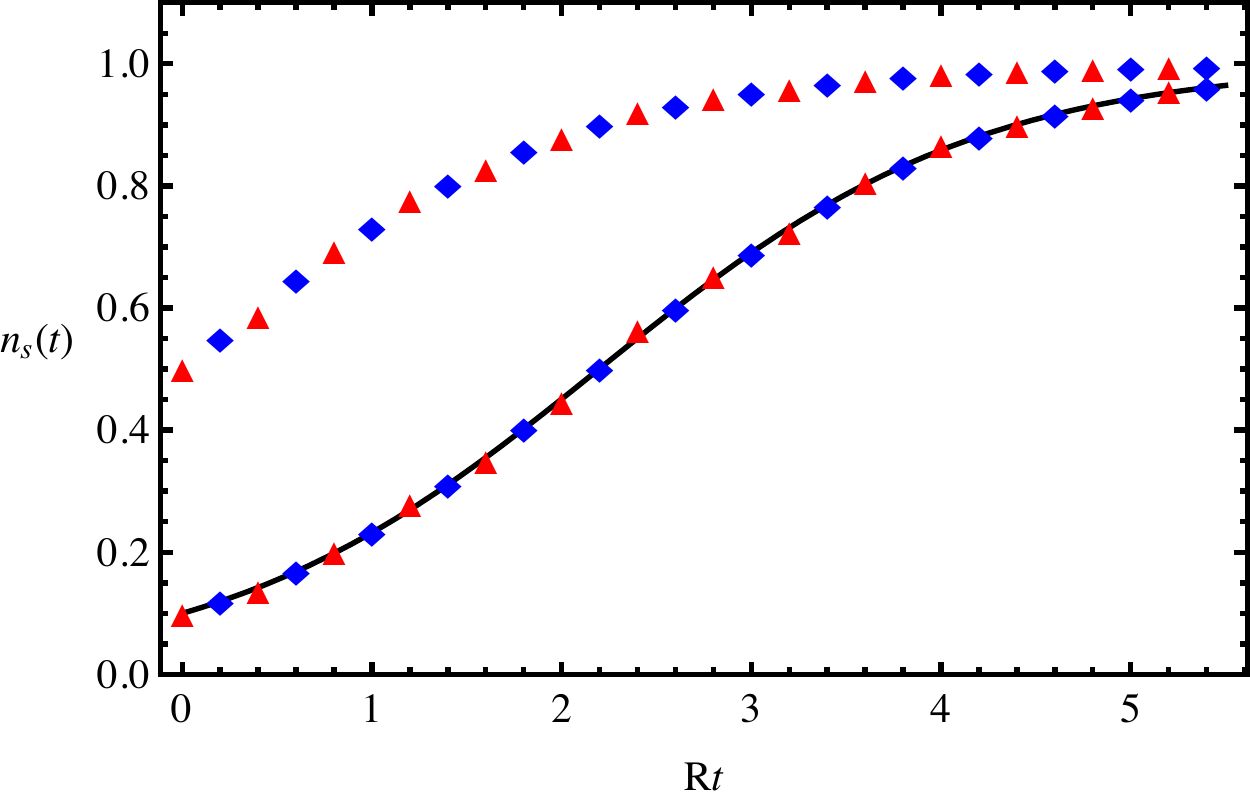,width=8.0cm}
\caption{(Color online) The signal population,  $n_s(t)$, plotted against $Rt$, revealing the universality of the QCM population-dynamics. The upper solid line corresponds to Eq. (\ref{nst}) for the case $n_1(0)=n_s(0)=0.5$, with the lower solid line showing the case $n_1(0)=0.9$, $n_s(0)=0.1$. The data points show the numerical results of the full QCM equations, (\ref{dc1dtau})-(\ref{dcsdtau}). The (blue) diamonds correspond to the Stark-regime, with $\Delta=10\Gamma D$, $\Omega=\Gamma D$, and the (red) triangles show the Rabi-regime, with $\Delta=0$ and $\Omega=10\Gamma D$.
\label{populationfig}}
\hspace{1cm}
\epsfig{figure=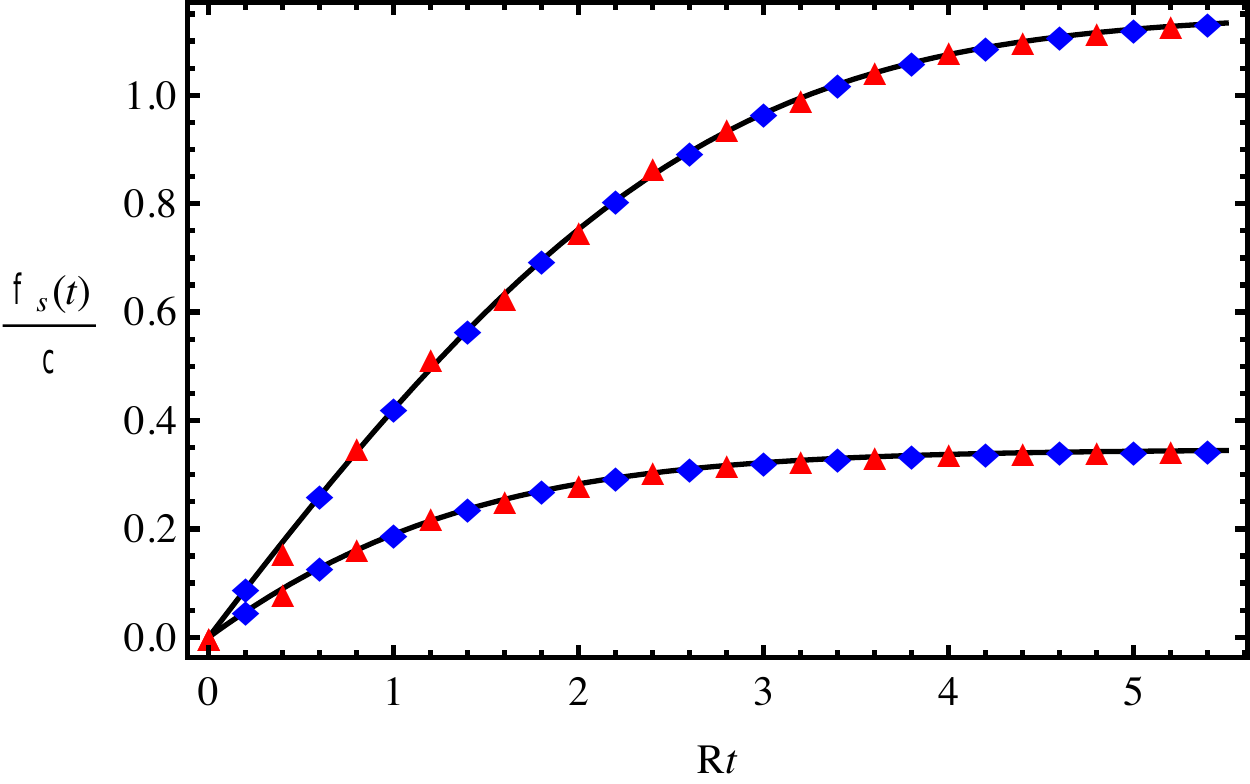,width=8.0cm}
\caption{(Color online) The re-scaled signal-mode phase,  $\phi_s(t)/\chi$, plotted against $Rt$, revealing the universality of the QCM phase-dynamics. The upper solid line corresponds to Eq. (\ref{phist}) for the case $n_1(0)=n_s(0)=0.5$, with the lower solid line showing the case $n_1(0)=0.9$, $n_s(0)=0.1$. The data points show the numerical results of the full QCM equations, (\ref{dc1dtau})-(\ref{dcsdtau}). The (blue) diamonds correspond to the Stark-regime, with $\Delta=10\Gamma D$, $\Omega=\Gamma D$, and the (red) triangles show the Rabi-regime, with $\Delta=0$ and $\Omega=10\Gamma D$.
\label{phasefig}}
\end{figure}

In figure \ref{populationfig} we plot the signal population $n_s(t)$ versus $Rt$, where $R$ is given by Eq. (\ref{R}). The solid lines show the universal analytic curve (\ref{nst}) for the two initial conditions, and the data points show numerical results corresponding to Stark, Zeno, and Rabi regimes. In figure \ref{phasefig}, we plot the signal-mode phase-shift acquired during MWA, comparing our analytic result to the exact numerical solution of Eqs. (\ref{dc1dtau})-(\ref{dcsdtau}). The figure plots $\phi_s(t)/\chi$, which normalizes out the dependence on the Fresnel number, ${\cal F}$, versus $Rt$, with $R$ again given by (\ref{R}). The solid lines show the universal analytic curve (\ref{phist}) for the two initial conditions, with the data points again showing numerical results corresponding to Stark, Zeno, and Rabi regimes.

In conclusion, for a given MWA setup, the fastest amplification of the signal wave is achieved in the Rabi regime, using an strong, resonant driving laser. Within the QCM approximation, both the population- and phase-dynamics is universal across the three main operating regimes. The population dynamics depends on geometry through the optical depth, $D=f_RN$, and is therefore optimized for speed by taking ${\cal F}=3.18$. The MWA phase-shift approaches a limiting value that scales linearly with $\chi=f_I/f_R$, and is thus minimized by taking ${\cal F}$ as large as possible. In this study that corresponds to a spherical condensate, as we have not considered the case of a pancake BEC geometry. For $v=10^7$, going from ${\cal F}=3.17$ ($A=23$) to ${\cal F}=210$ ($A=1.0$) results in a decrease in the MWA gain rate $R$ by a factor of $5$, with a decrease in the phase-shift by a factor $45$.

\subsubsection{Mode-competition and the MWA threshold}
Atom losses in the QCM model are described  by the $\tilde\gamma$ term in Eq. (\ref{dcedtau}).  These losses correspond to the depletion of the atomic mean-field associated with the spontaneous emission of incoherent radiation. In this section, we consider the possibility that some spontaneous photons are emitted into modes that, while orthogonal to the idler mode, still lie within the solid-angle for end-fire emission, and therefore have a large gain-factor for amplification. Such amplified spontaneous emission is typically referred to in this situation as matter-wave superradiance \cite{InoChiSta99}. Here we include atom losses in our analytical approximations, after scaling time to the MWA rate constant, $R$, the superradiant decay of the pump condensate is only weakly dependent on the number of end-fire modes. Furthermore, we show that the effect of pump depletion due to incoherent emission depends solely on the optical depth.

The adiabatic elimination of the excited state in the Stark regime, and the introduction of dressed-states in the Rabi regime lead to two seemingly incongruous  mean-field theories, given by Eqs. (\ref{dc1dtauStark}-\ref{dcsdtauStark}) and (\ref{dcpmdtauRabi}-\ref{dcsdtauRabi}). With the introduction of the pump population, $N_p$, given in the Stark regime by $N_p=N_1$ and in the Rabi regime by $N_p=N_++N_-$, i.e. the net population of the two dressed states defined by \ref{cpm}, the population dynamics of the pump and signal modes obey a single set of rate equations,
\beqa
\ddt N_p &=&-\frac{R}{N}N_p N_s\\
\ddt N_s&=&\frac{R}{N}N_pN_s.
\eeqa
These equations can be generalized to include incoherent emission and competing superradiance via the addition of spontaneous emission terms, and the inclusion of a complete set of atomic field modes, $\{|\bfk,j \ra\}$, where $\bfk$ labels the spatial mode, and $j$ the internal hyperfine level, which yields
\beqa
&&\!\!\!\!\!\!\!\!\ddt N_p=-\frac{R}{N}N_p\left(\!\!N_s+1+\sum_{\bfk j}s_{\bfk j}\left(N_{\bfk j}{+}1\right)\!\!\right)\!\!,\label{ddtNp1}\\
&&\!\!\!\!\!\!\!\!\ddt N_s=\frac{R}{N} N_p(N_s+1),\\
&&\!\!\!\!\!\!\!\!\ddt N_{\bfk j}=\frac{R}{N}N_ps_{\bfk j}(N_{\bfk j}+1),
\eeqa
where 
\beq
s_{\bfk j}=\frac{\Gamma_j}{\Gamma}\frac{f_{kj}}{f_R},
\eeq
is the ratio  ratio of the gain-factor for the $|j,\bfk\ra$ mode to that of the signal mode, satisfying 
\beq
1+\sum_{\bfk j}s_{\bfk j}=\frac{\gamma}{\Gamma}\frac{N}{D}.
\eeq
For the level schemes  depicted in figures \ref{scheme} and \ref{systems}, because the upper and lower levels are both $J=1$ levels, decay to the $m_J=0$ ground state is forbidden. Thus there are two decay channels with $\Gamma_1=\Gamma_2=\Gamma$, so that $\gamma=2\Gamma$. Henceforth we shall assume these conditions hold.
Introducing the fractional populations $n_p$, $n_s$, and $n_{\bfk j}$, and scaling time by the rate constant $R$,  we arrive at
\beqa
&&\!\!\!\!\!\!\!\!\!\!\!\!\ddt n_p = \!\!{-}n_p\!\left(n_s+\sum_{\bfk j}s_{\bfk j}n_{\bfk j}+\frac{2}{ D}\!\right)\!\!,
    \label{mmmt1}\\
     &&\!\!\!\!\!\!\!\!\!\!\!\!\ddt n_s= n_p \left(n_s+\frac{1}{N}\right),\label{mmmt2}\\
 &&\!\!\!\!\!\!\!\!\!\!\!\!\ddt n_{\bfk j}= n_p s_{\bfk j} \left(n_{\bfk j}+\frac{1}{N}\right).\label{mmmt3}
 \eeqa
 We note that the incoherent loss rate in (\ref{mmmt1}) of $2/D$, derived from these ad-hoc rate equations is exactly the rate derived from the full quantum-field theory (\ref{dcedtau}), $\tilde\gamma = 2/D$. 
 
For a typical elliptical condensate with $\cA\gg 1$ we have
\begin{equation}
\label{skj}
 s_{\bfk j}\approx\frac{1-|\hat{d}_j|^2)}{\sqrt{\cos^2\theta_\bfk+\cA^2\sin^2\theta_\bfk}},
\end{equation}
where $\theta_\bfk$ is angle between $\bfk$ and $\hat{z}$. The denominator gives the effect of geometry, whereas the numerator gives the dipole-emission pattern.  For condensates with a large aspect ratio, $\cA\gg 1$, the transition to a strongly suppressed gain occurs at the end-fire angle $\theta_{EF}=W/L$. Thus, all side modes can be classified into end-fire (EF) modes, corresponding to $\theta_\bfk<\theta_{EF}$, and non-end-fire (NEF) modes, corresponding to $\theta_\bfk>\theta_{EF}$. We will assume that a given NEF mode never builds up any population, but that EF modes may become macroscopically populated due to superradiance. To greatly simplify the analytics,  we can therefore approximate $s_{\bfk j}$ by a step function
\begin{equation}
\label{fkj}
    s_{\bfk j}\approx  u(\theta_{EF}-\theta_\bfk).
\end{equation}
Using the conservation law $n_p+n_s+\sum_{\bfk j}n_{\bfk j}=1$ to eliminate $n_s(t)$, results in a closed equation for $n_p$,
\beq
\ddt n_p=-n_p\left(1-n_p+\frac{2}{D}\right),
\eeq
 which has the solution
 \beq
 n_p(t)=\frac{n_p(0)\left(1+\frac{2}{D}\right)}{n_p(0)+\left(1+\frac{2}{D}-n_p(0)\right)e^{Rt(1+2/D)}}.
 \eeq
In the limit $D\ll 1$, this describes ordinary exponential decay,
\beq
\lim_{D\to 0} n_p(t)=n_p(0)e^{-2Rt/D},
\eeq
while in the opposite limit of a large optical depth, it describes super-exponential decay
\beq
\lim_{D\to\infty} n_p(t)=\frac{n_p(0)e^{-Rt}}{1+n_p(0)(e^{-Rt}-1)}.
\eeq

More important to consider is the signal mode population, given in the limit $N\gg 1$ by
\beq
n_s(t)=\frac{n_s(0)\left(1+\frac{2}{D}\right)}{n_s(0)+\frac{2}{D}+(1-n_s(0))e^{-Rt(1+2/D)}}.
\eeq
To see the existence of an MWA threshold, consider the final population of the signal mode,
\beq
n_s(\infty)=\frac{n_s(0)\left(1+\frac{2}{D}\right)}{n_s(0)+\frac{2}{D}}.
\eeq
In the limit $D\to 0$, this gives $\lim_{D\to0}n_s(\infty)=n_s(0)$, so that no MWA has occurred. In the opposite limit $D\to\infty$, it gives
$\lim_{D\to\infty}n_s(\infty)=1$, which is the result with no competing incoherent losses. 

The threshold for MWA can therefore be defined as $n_s(\infty)=0.5$, which means that lossless amplification of an initial signal of $N_s$ using a pump condensate 
of $N_p$ atoms  requires an optical depth of $D\gg D_c$, where
\beq
D_c=2\left(\frac{N_p}{N_s}-1\right).
\eeq
In our numerical simulations, $N_p(0)=9\times 10^5$, $N_s(0)=1\times 10^5$, so that $D_c=16$. Choosing $\cF=3.18$ to maximize the optical depth results in $D=1050$, while a typical aspect of $\cA=10$ results in $D=853$, so that losses are essentially negligible during the MWA process.  From a different perspective, we can conclude that lossless amplification of a signal condensate of $N_s$ atoms using a pump condensate of $N_p$ atoms having optical depth $D$ requires
\beq
N_s\gg \frac{2 N_p}{2+D}.
\eeq
For $N_p\sim 10^6$ and $D\sim 10^3$,  the initial signal population must therefore satisfy $N_s\gg 10^3$, i.e. $n_s(0)\gg 10^{-3}$, for strong amplification to occur. 
The final occupation of a typical competing EF mode is given for $D\gg 1$ by $N_{EF}(\infty)=N_p/N_s>1$. 
For the parameters of our simulations there are $M\sim \cF^2$ EF modes, so the total number of atoms transferred to competing modes for the case $\cF=3.18$ is $M N_{EF}\sim 10^2$, which is $0.01\% $ of the total atoms. For comparison, about $2\%$ of the total atoms are lost ($n_{loss}=1-n_s(\infty)$), the vast majority of which are emitted into the NEF modes, which vastly outnumber the EF modes.

\section{Spatial dependence of the matter-wave dynamics}
\label{SPE}

The primary limitation of the QCM model is not that it neglects the spatial
dependence of the light fields inside the condensates, e.g. the idler field is zero at the front of the BEC, and strongest towards the back. The limitation is  that it does not allow for a similar spatial dependence of thewavefunction of the condensate, which in the case of Rayleigh MWA would correspond to a spatial modulation in the visibility of the matter-wave density grating. In practice,
such effects have been shown to account for several important
experimental observations \cite{AveTri04,
ZobNik05,ZobNik06,UysMey07}. 

To incorporate these spatial propagation effects, we now 
turn to numerical simulation of the mean-field Equations (\ref{ptpsi1})-(\ref{ptpsi3}).
Assuming a cylindrically symmetric system, we can make the change of variables
\beqa
\psi_1(\bfr)&=&\sqrt{N}\tilde\psi_1(\rho,z), \\
\psi_2(\bfr)&=&\sqrt{N}\tilde\psi_s(\rho,z),\\
\psi_3(\bfr)&=&\sqrt{N}\tilde\psi_e(\rho,z)e^{ik_Lz},
\eeqa
 the mean-field equations become
\begin{widetext}
\beqa
&&\!\!\!\!\!\!\!\!\!\!\!\!\pptau\tpsi_1(\rho,z)=\frac{i}{2}\tilde\Omega\str\tpsi_e(\rho,z)
+\frac{i}{2}\int \rho'd\rho'\,dz'\, \tG\str(\rho,z;\rho',z')\tpsi_e\str(\rho',z')\tpsi_1(\rho',z')\tpsi_e(\rho,z,\tau),\label{tpsi1}\\
&&\!\!\!\!\!\!\!\!\!\!\!\!\pptau\tpsi_s(\rho,z)=\frac{i}{2}\int \rho'd\rho'\,dz'\,\tG\str(\rho,z;\rho',z')\tpsi_e\str(\rho',z')\tpsi_s(\rho',z')
\tpsi_e(\rho,z),\label{tpsis}\\
&&\!\!\!\!\!\!\!\!\!\!\!\!\pptau\tpsi_e(\rho,z)=\left[-i\tDelta-\frac{\tgamma}{2}\right]\tpsi_e(\rho,z)+\frac{i}{2} \tOmega\tpsi_1(\rho,z)
+\frac{i}{2}\sum_{j=1,s}\int \rho'd\rho'\,dz'\,\tG(\rho,z;\rho',z')\tpsi_j\str(\rho',z')\tpsi_e(\rho',z')\tpsi_j(\rho,z),\label{tpsie}
\eeqa
with
\beq
\tG(\rho,z;\rho',z')=\frac{3}{4f_R}\int_0^{2\pi} d\phi'\, \left[1+\frac{k_L^2(z-z')^2}{\xi^2(\phi')}
+\left(3\frac{k_L^2(z-z')^2}{\xi^2(\phi')}-1\right)\left[\frac{i}{\xi(\phi')}-\frac{1}{\xi^2(\phi')}\right]
\right]\frac{e^{i[\xi(\phi')-k_L(z-z')]}}{\xi(\phi')},\label{tG}
\eeq
\end{widetext}
where 
\beq
\xi(\phi')=k_L\sqrt{(z-z')^2+\rho^2+\rho'{^2}-2\rho\rho'\cos\phi'},\label{xi}
\eeq
and as with the QCM model we have $\tau=\Gamma D t$, $\tOmega=\Omega/\Gamma D$, $\tDelta=\Delta/\Gamma D$, with $D=f_RN$.

\subsection{Adiabatic elimination in the Stark regime}
For the Stark regime, $\tDelta^2\gg |\tOmega|^2+1$, we can adiabatically eliminate the excited state dynamics by expanding $t\psi_e$ to second-order in powers of $1/\tDelta$. This gives 
\beq
\tpsi_e\approx \frac{\tOmega}{4\tDelta^2}
\left(i\tgamma+2\tDelta++\tG[\tpsi\str_1\tpsi_1]+\tG[\tpsi\str_s\tpsi_1]\right)\tpsi_1,
\eeq
which leads to
\beqa
\pptau\tpsi_1&=&\frac{i}{2}\frac{|\tOmega|^2}{4\tDelta^2}
\left(2\tDelta+i\tgamma+(\tG+\tG\str)[\tpsi_1\str\tpsi_1]\right)\tpsi_1\nn
&+&\frac{i}{2}\frac{|\tOmega|^2}{4\tDelta^2}\tG[\tpsi_s\str\tpsi_1]\tpsi_s\label{tpsi1eq}\\
\pptau\tpsi_e&=&\frac{i}{2}\frac{|\tOmega|^2}{4\tDelta^2}\tG\str[\tpsi_1\str\tpsi_s]\tpsi_1.\label{tpsiseq}
\eeqa
The first term in Eq. (\ref{tpsi1eq}) describes the decay of the mean-field due to spontaneous emission. The second term is the usual spatially homogeneous AC-Stark shift. with the third term describing the spatial-variation to the AC-Stark shift due to any modulation of the pump laser amplitude caused by the atom-field interaction. The remaining terms in Eqs. (\ref{tpsi1eq}) and (\ref{tpsiseq}) describe the transfer of atoms from the pump-mode to the signal mode which lies at the heart of the MWA process.

With the change of variables, $\tau=\frac{4\tDelta^2}{|\tOmega_L|^2}|\tau'$ and $\tpsi_1=\tpsi_1'e^{i\tDelta\tau'}$, these simplify to
\beqa
\partial_{\tau'}\tpsi_1'&=&-\frac{\tgamma}{2}\tpsi_1'+\frac{i}{2}(\tG+\tG\str)[\tpsi_1'{\str}\tpsi_1']\tpsi_1'\nn
&+&\frac{i}{2}G[\tpsi_s\str\tpsi_1']\tpsi_s\label{adelimpsi1}\\
\partial_{\tau'}\tpsi_s&=&\frac{i}{2}G\str[\tpsi_1'{\str}\tpsi_s]\tpsi_1'.\label{adelimpsis}
\eeqa
Note that these equations are invariant under the transformation $\tpsi_s\to e^{-i\phi}\tpsi_s$, which means that the population and phase-shift dynamics are independent of any spatially uniform initial relative phase between $\psi_1$ and $\psi_s$.
This shows that in the Stark-regime, the signal mode population and phase dynamics is unaffected by the spatially homogeneous part of the AC-Stark shift, and its dependence on the pump intensity and detuning disappears upon scaling time to the rate constant $R=\frac{|\Omega_0|^2\Gamma D}{4\Delta^2}$, just as with the QCM model.

\subsection{Numerical simulations}
In this section we give results of numerical simulations of the mean-field theory given by Eqs. (\ref{tpsi1})-(\ref{tpsie}). We will compare results using the full Green's function of Eq. (\ref{tG}), to those using the Paraxial Green's function, Eq.  (\ref{GP}), and the Longitudinal Green's function, Eq. (\ref{GL}) as well as the results of the QCM model, Eqs. (\ref{dc1dtau})-(\ref{dcedtau}). To study the limit of very large detuning, we instead solve the adiabatic equations, (\ref{adelimpsi1}) and (\ref{adelimpsis}). The spatial mean-field equations are solved using a grid of $20$ points along the radial direction and $200$ points along the longitudinal axis. This requires first computing and storing the greens function as a $4000\times 4000$ array, which prevents us from increasing the number of mesh points due to computer memory limitations. One test of the adequacy of this grid is to use it to compute numerically the collectivity parameters $f_R$ and $f_I$. We find that for the full Green's function of Eq. (\ref{tG}), the deviation from the exact values of $f_R$ and $f_I$ is $1\%$. Based on this, we conclude that the $20\times200$ grid is sufficient to accurately model the true electric field inside the condensate during the MWA process.

We consider a condensate of $N=10^6$ atoms confined to a dimensionless volume of $v=10^7$. For initial conditions, we take $\tpsi_1(\bfr,0)=\sqrt{n_1(0)}\phi(\bfr)$, $\tpsi_s(\bfr,0)=\sqrt{n_s(0)}\phi(\bfr)$, and $\tpsi_e(\bfr,t)=0$, with $n_s(0)$ being the initial signal population fraction, and $n_1(0)$ being the initial pump fraction, so that $n_1(0)+n_s(0)=1$, and with the mode function $\phi(\bfr)$ given by the Thomas-Fermi wavefunction of Eq. (\ref{TF}). The pump laser is taken to be spatially uniform, with a sudden turn-on at $t=0$.

\subsubsection{Stark Regime}
\begin{figure}
\epsfig{figure=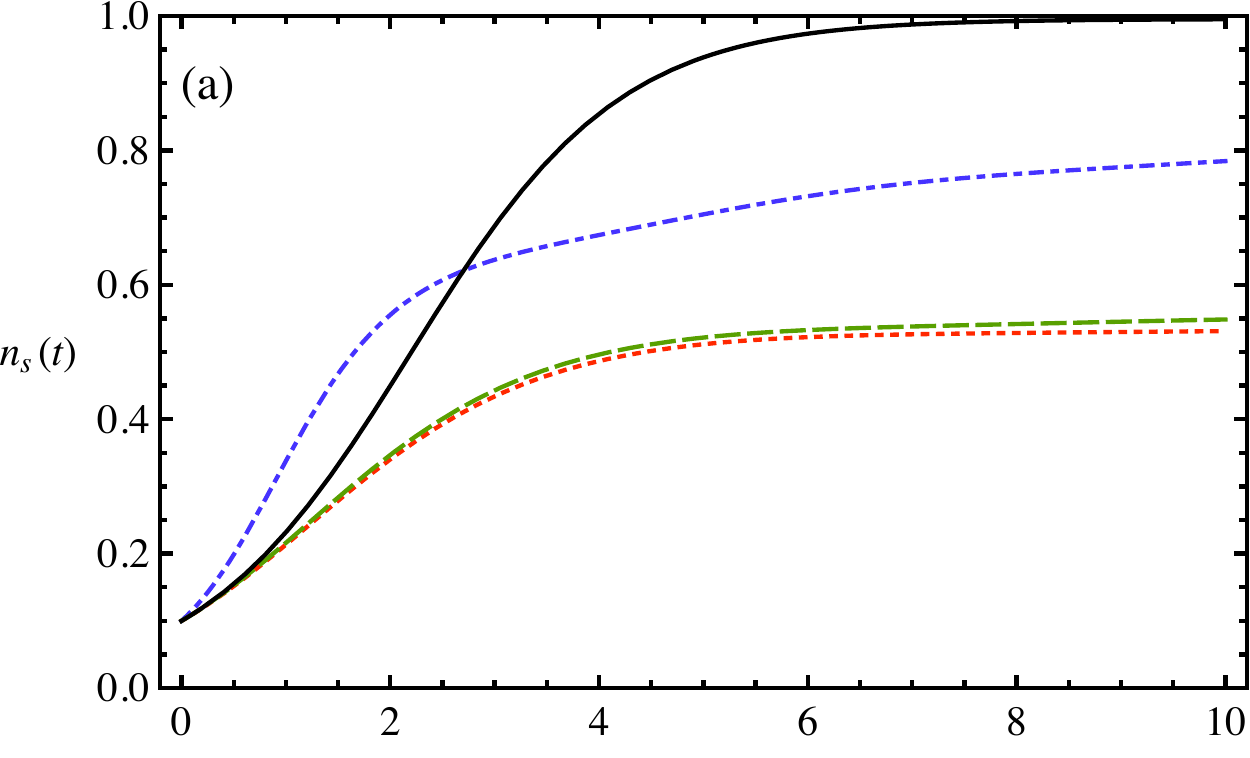,width=8.0cm}
\epsfig{figure=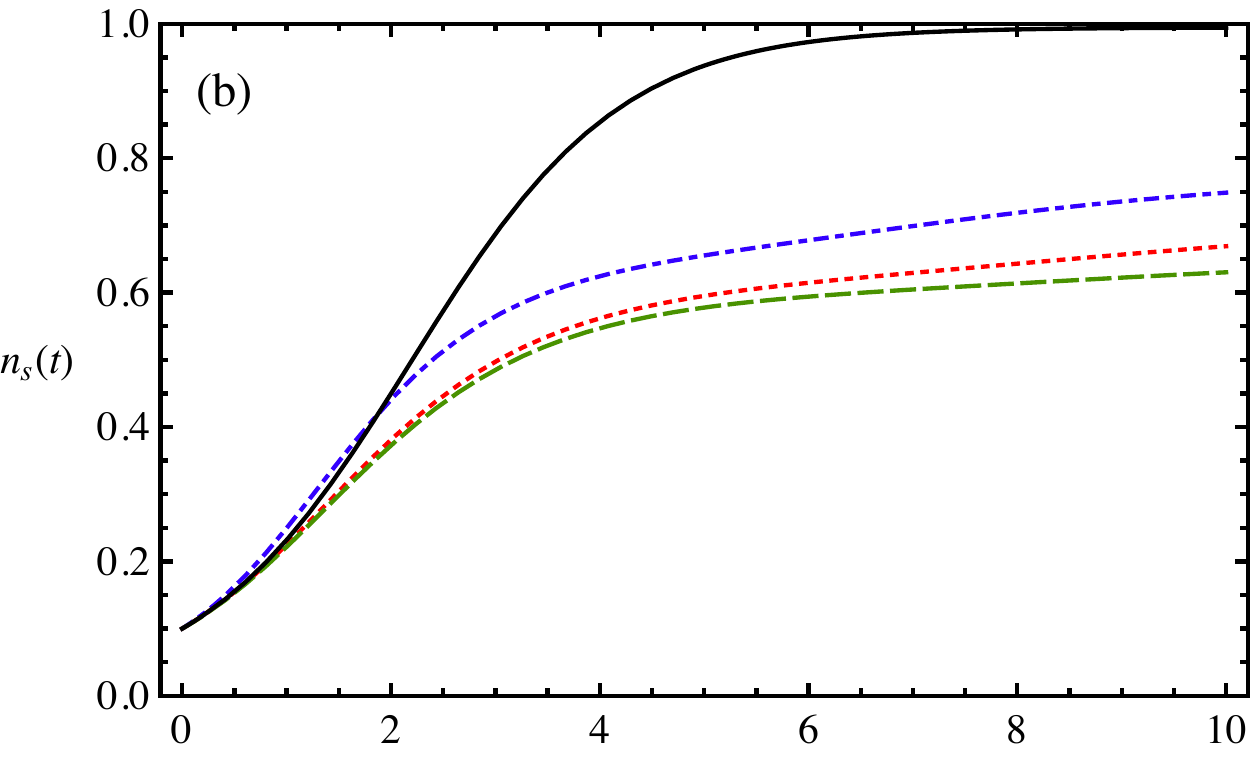,width=8.0cm}
\epsfig{figure=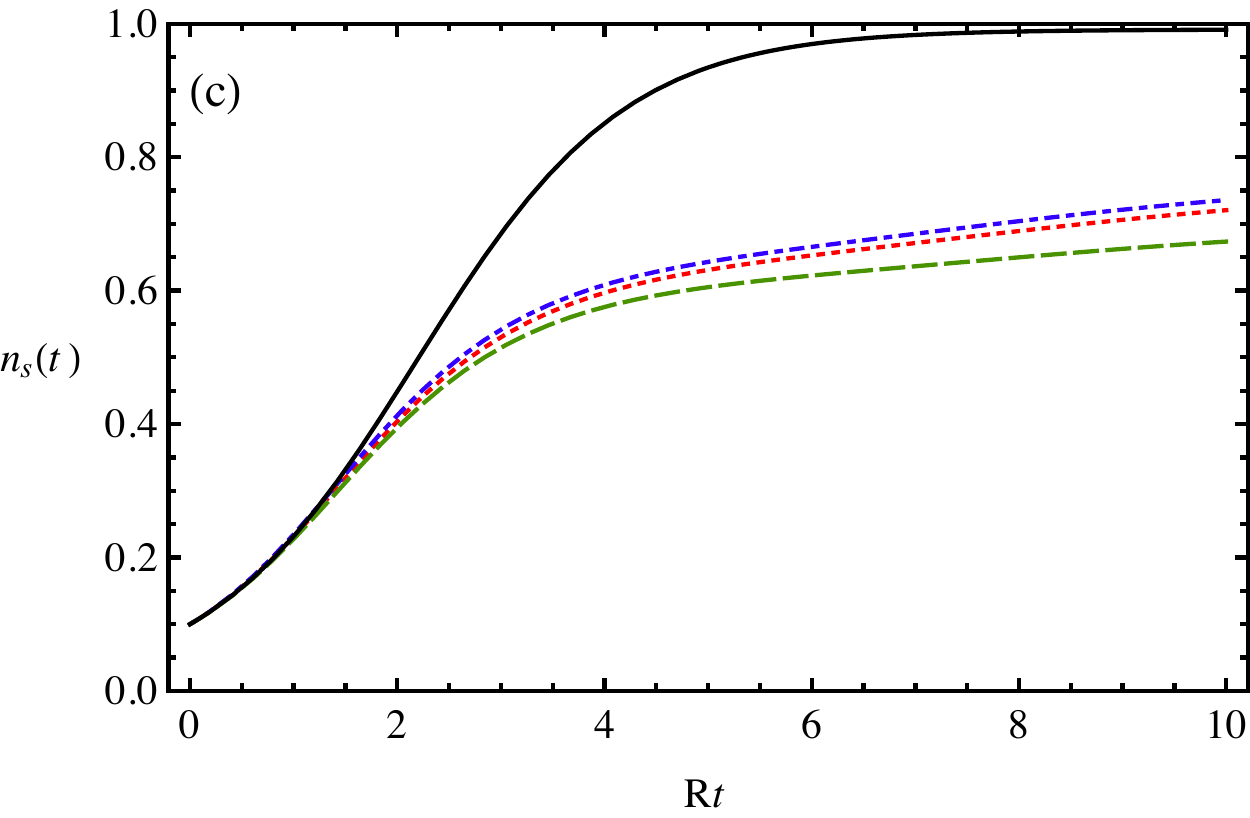,width=8.0cm}
\caption{(Color online) The signal mode population fraction $n_s(t)$ versus $Rt$, for the Stark regime, determined by Eqs. (\ref{adelimpsi1}) and (\ref{adelimpsis}), for the initial conditions $n_s=0.1$ and $n_1=0.9$, with $N=10^6$ and $v=10^7$. In all figures, the (black) solid line shows the results of the QCM model, the (red) dotted line shows the results for the full Green's function, the (green) dashed line shows the results for the paraxial Green's function, and the (blue) dash-dotted line shows the results for the Longitudinal Green's function. Figure (a) shows the  case $\cF=3.18$, corresponding to an aspect ratio of $\cA=24$, figure (b) shows the case $\cF=10.0$, corresponding to an aspect ratio of $\cA=10$, and figure (c) shows the case $\cF=30$, corresponding to $\cA=4.4$.
\label{Starkns91}}
\end{figure}

For the off-resonant Stark regime, defined as $\Delta^2\gg \Omega^2+\Gamma^2D^2$, we have numerically solved the adiabatic equations (\ref{adelimpsi1}) and (\ref{adelimpsis}) for a variety of geometries and initial conditions.  In Figure \ref{Starkns91} we  the signal-mode population fraction, $n_s(t)$ versus $Rt$, where 
\beq
n_s(t)=\int d\bfr |\tilde\psi(\bfr_\perp,z)|^2,
\eeq 
and $R$ is given by Eq. (\ref{R}). The initial conditions are taken as $n_s(0)=0.1$ and $n_1(0)=0.9$, corresponding to a small-signal amplification process.  The solid (black) line gives the results from the QCM (single-mode) model, the dash-dotted line (blue) shows the results from the longitudinal Green's function, the dashed (green) line shows the results from the Paraxial Green's function, and the dotted (red) line shows the results from the exact Green's function.

In Figure \ref{Starkns91}a, we show the results for $\cF=3.18$, corresponding to an aspect ratio of $\cA=24$, which gives the maximum gain within the QCM model. In this case, we see that there is excellent agreement between the full and Paraxial Green's functions, which is expected for the case of a large aspect ratio. We also see that there is exact agreement between the single-mode approximation (QCM) and the multi-mode exact model at short times
At longer times, the exact model saturates with only about half of the pump atoms transferred to the signal mode, whereas the single-mode theory predicts a complete transfer. The results from the longitudinal Green's function do not agree with either the QCM or exact models, greatly overestimating the initial gain and final saturation level.

In Figure \ref{Starkns91}b, we depict the case $\cF=10$, corresponding to a shorter aspect ratio of $\cA=10$, which is typically of many MWA and superradiance experiments. Here we see that for short times there is again agreement between the 
single-mode QCM model and the exact model. At longer times, the paraxial approximation is not quite as good as with the $\cA=24$ case. The inaccuracy of the longitudinal model is slightly improved. The final saturation level of the signal-mode population is seen to increase with respect to the $\cF=3.18$ case. 

Lastly, in Figure \ref{Starkns91}c, we show results for $\cF=30$, corresponding to an even shorter aspect ratio of $\cA=4.4$.
In this case, we now see good agreement between the Longitudinal and exact Green's functions. Again the agreement with the QCM model is exact at short times, followed by a break-away and transition to a very slow rate of increase in the signal-mode population.

\begin{figure}
\epsfig{figure=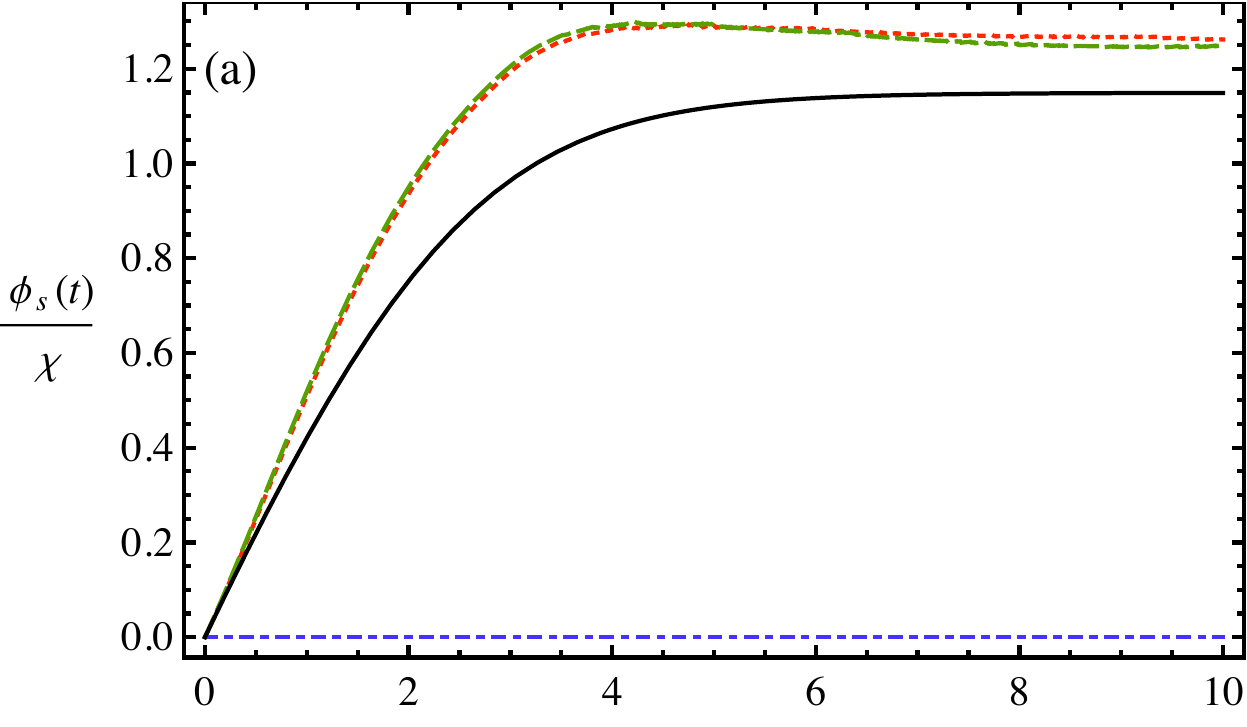,width=8.0cm}
\epsfig{figure=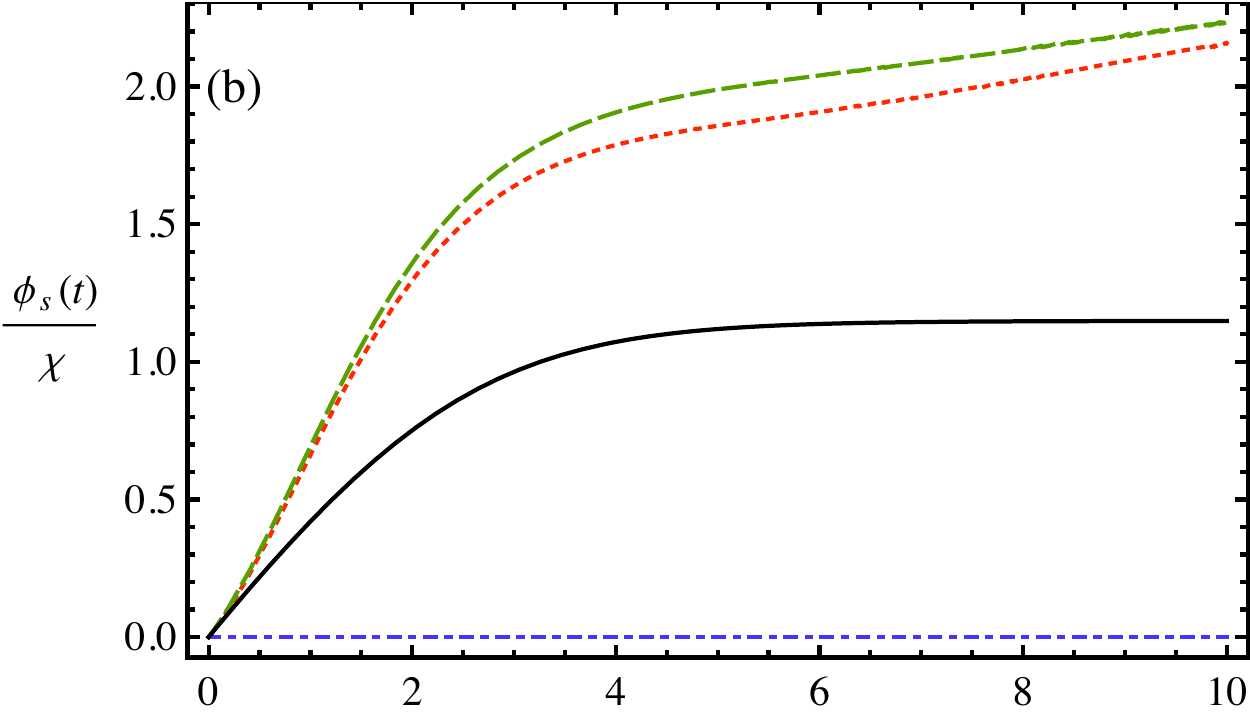,width=8.0cm}
\epsfig{figure=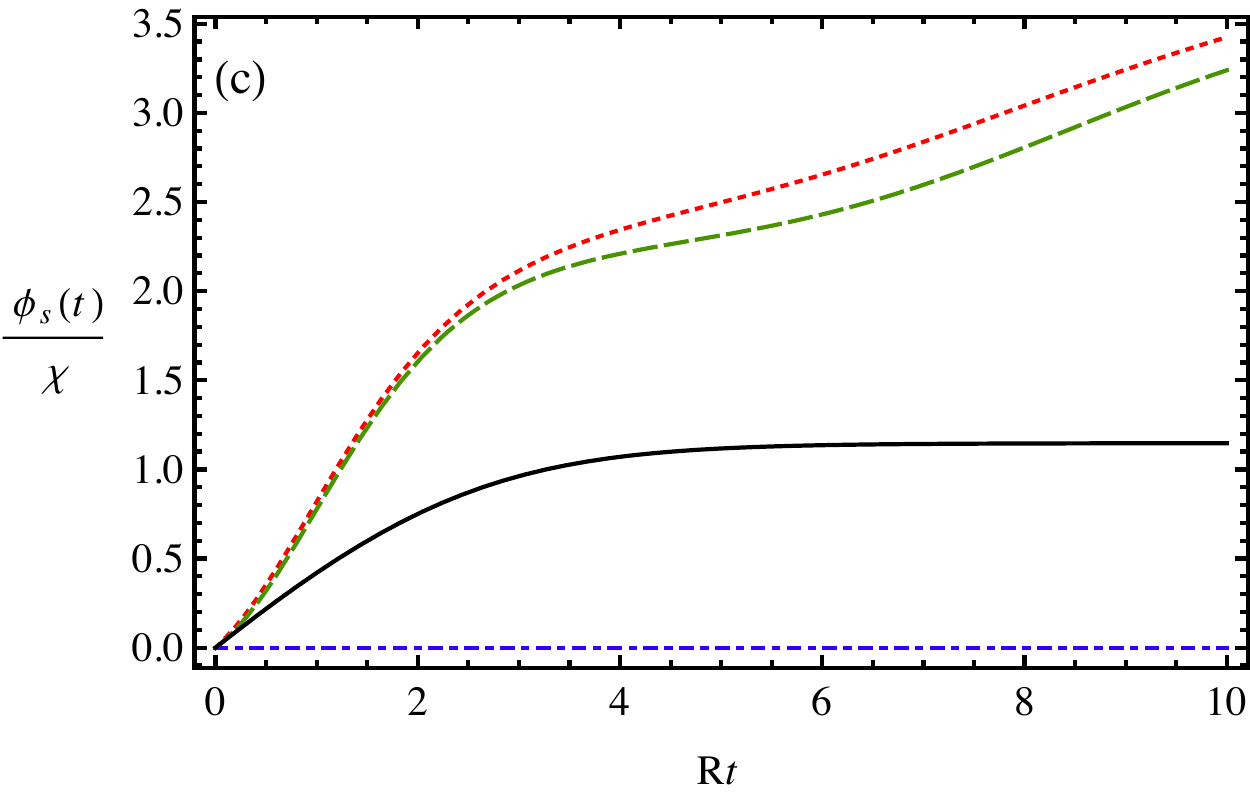,width=8.0cm}
\caption{(Color online) The spatially averaged signal-mode phase, $\phi_s(t)$ scaled by $\chi$ and  plotted versus $Rt$ for the Stark regime. All parameters are as in Figure \ref{Starkns91}.
\label{Starkphis91}}
\end{figure}

We now turn our attention to the phase-dynamics of the signal matter-wave. Ideally, a MWA device should leave the signal-mode phase unaltered. From the QCM model, we expect that this is not possible within the current paradigm. If the induced phase shift can be calculated precisely, however,  then it can be taken into account so that an MWA device can still be employed for interferometric applications.  In figure \ref{Starkphis91},  we plot the signal-mode phase divided by $\chi$ versus $Rt$ for the same parameters as in Figure \ref{Starkns91}.  The signal mode phase has been spatially averaged, weighted by the signal-mode density, according to
\beq
\phi_s(t)\equiv\frac{1}{n_s(t)}\int d\bfr |\psi_s(\bfr,t)|^2\arg[\phi_s(t)].
\eeq
From Fig. 8, we see that with respect to the phase dynamics, the agreement between the QCM model and the exact result, while still perfect at very short times, diverges much more quickly than the population dynamics, more so with increasing $\cF$. Thus the closest agreement with the QCM model is in the $\cF=3.18$ case. Note that the Longitudinal Green's function model predicts zero phase-shift in all cases. Thus even for the case of a large Fresnel number, where the Longitudinal model accurately predicts the population dynamics, as seen in Fig. \ref{Starkns91}c,  it does not accurately model the phase dynamics.

\subsubsection{Rabi Regime} 
\begin{figure}
\epsfig{figure=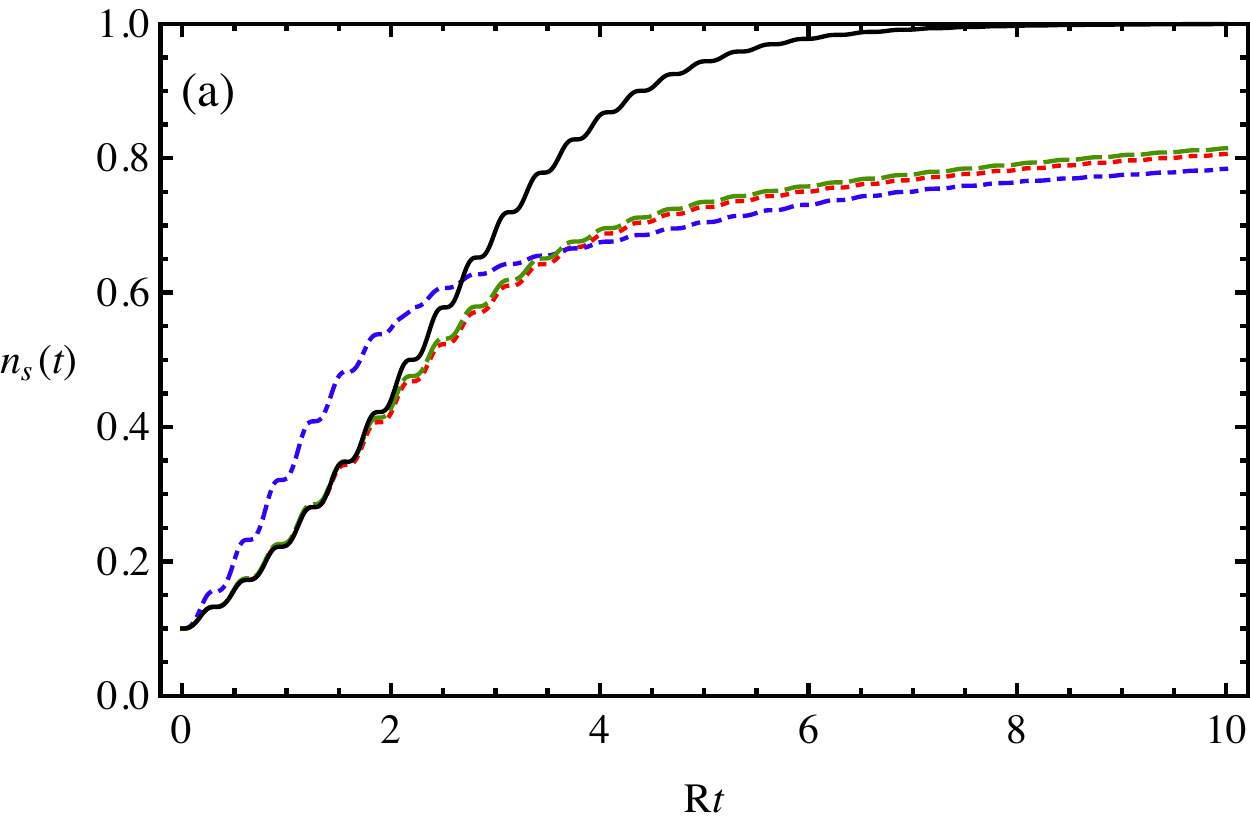,width=8.0cm}
\epsfig{figure=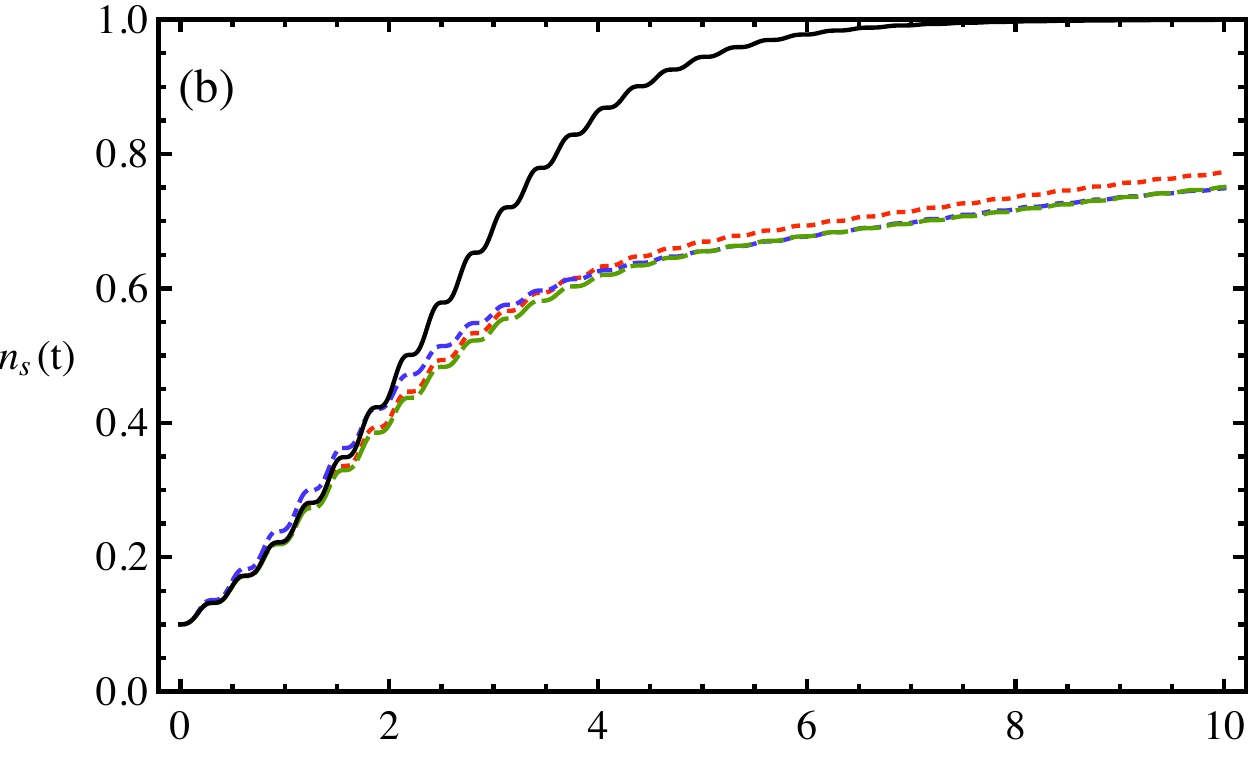,width=8.0cm}
\epsfig{figure=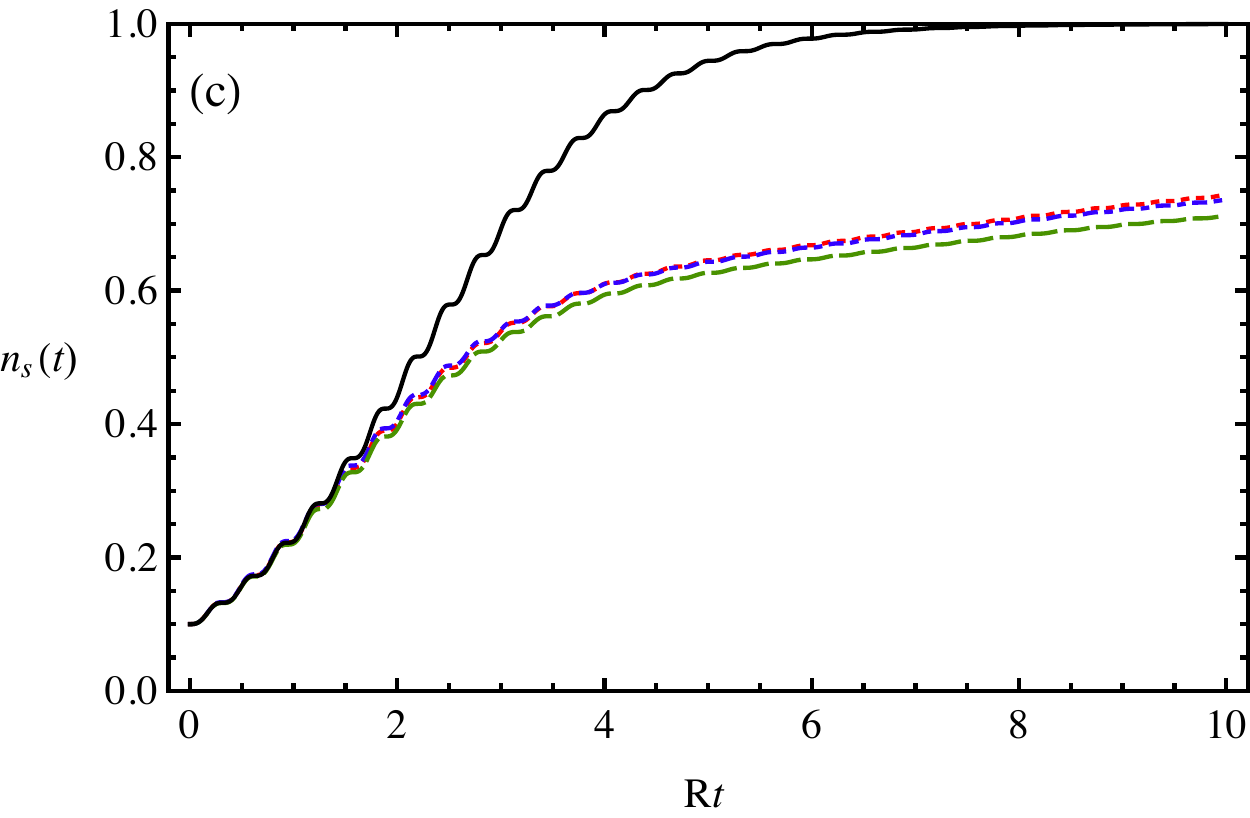,width=8.0cm}
\caption{(Color online) The signal mode population fraction, $n_s(t)$ versus $Rt$ for the Rabi regime, with $\Delta=0$ and $\Omega_L=10\Gamma D$. The initial conditions and geometric parameters are the same as in figure \ref{Starkns91}.
\label{Rabins91}}
\end{figure}

For the on-resonance Rabi regime, we take $\tilde\Delta=0$ and $\tilde\Omega=10$, and solve numerically Eqs. (\ref{tpsi1})-(\ref{tpsie}). The signal-mode population dynamics is poltted versus $Rt$ for initial conditions $n_s(0)=0.1$ and $n_1(0)=0.9$,  in Fig. \ref{Rabins91}. Figures\ref{Rabins91}a, b, and c correspond to $\cF=3.18$, $10$, and $30$, respectively.  As with the Stark regime, we see good agreement between the full Green's function and the Paraxial Green's function for the highly-elongated $\cF=3.18$ case, with large discrepancies occurring for larger Fresnel numbers. In the opposite limit of a relatively small aspect ratio, we see for $\cF=30$ that there is good agreement instead between the full Green's function and the Longitudinal approximation.

In figure \ref{Rabins91}a, we see a very important contrast between the on-resonant and the off-resonant dynamics in the maximum gain, $\cF=3.18$ case. Unlike the off-resonant case of Fig. \ref{Starkns91}a, where the signal mode population determined via the full Green's function saturates at $n_s(t)\sim .5$, well below the prediction of the longitudinal Green's function, in the on-resonant case, the saturated value of $n_s(t)$ is even slightly above that of the Longitudinal model. This effect is also seen in figures \ref{Rabins91}b and c, albeit the differences are less pronounced.

\begin{figure}
\epsfig{figure=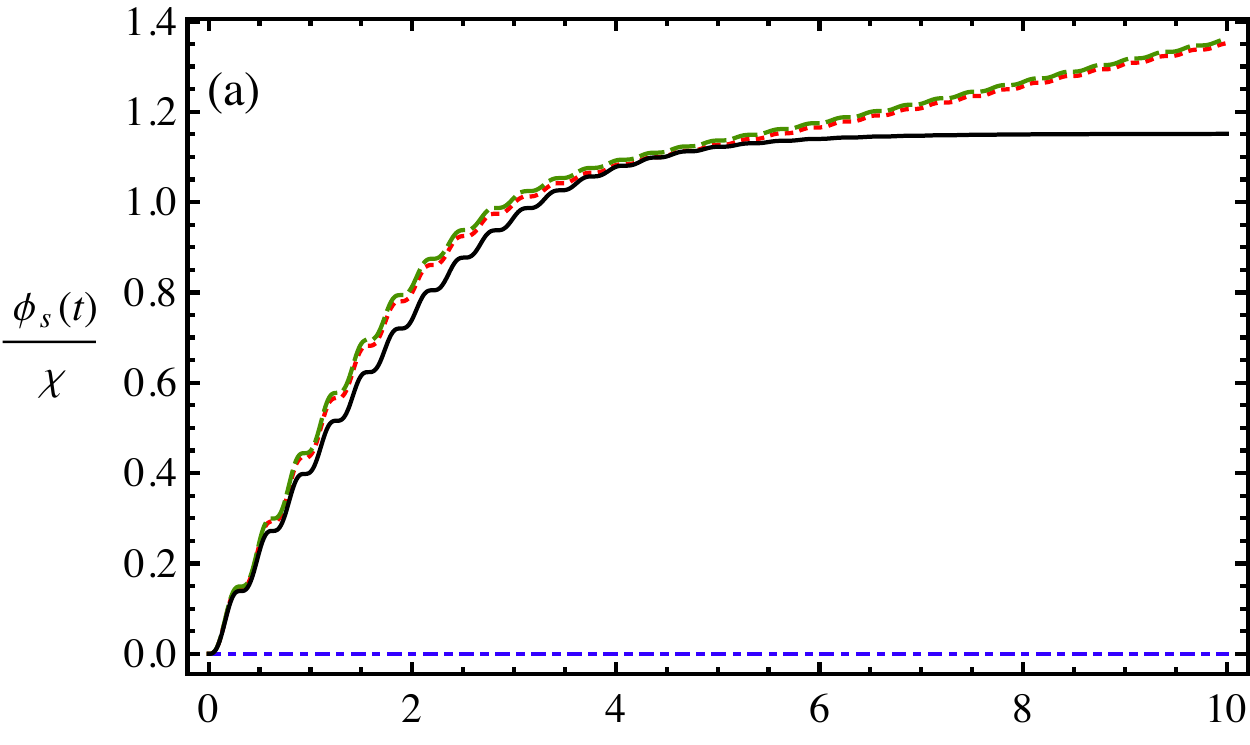,width=8.0cm}
\epsfig{figure=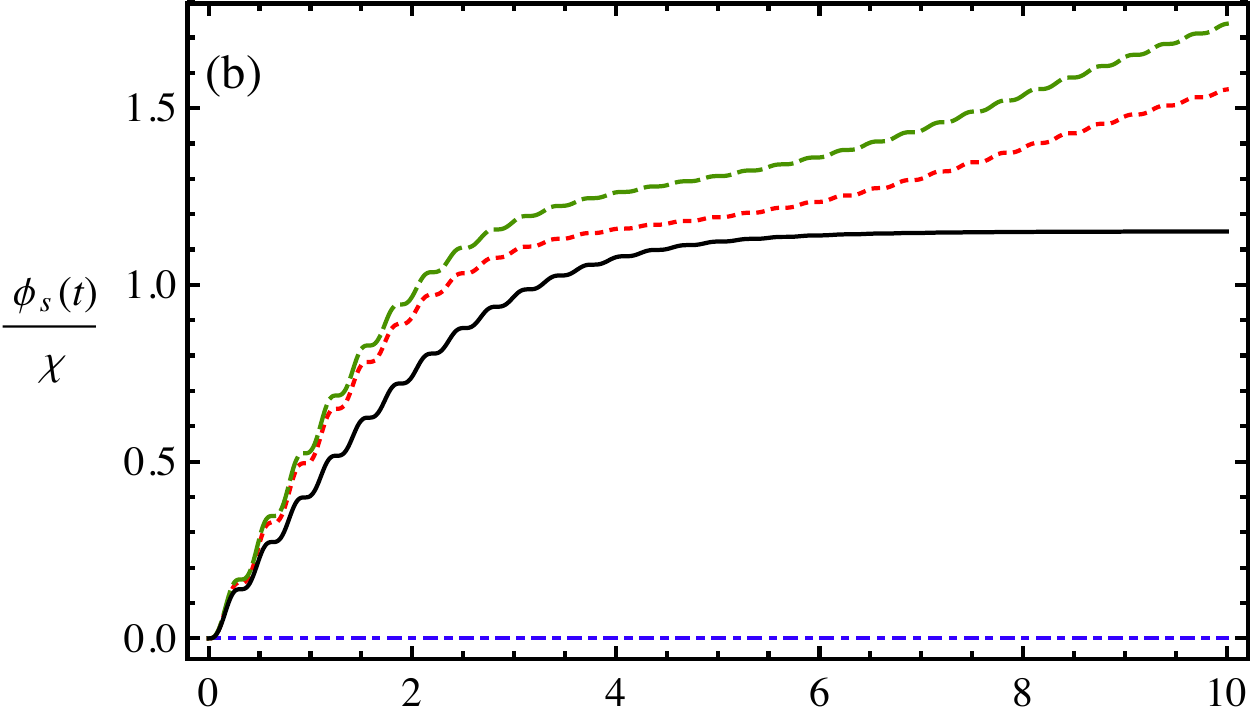,width=8.0cm}
\epsfig{figure=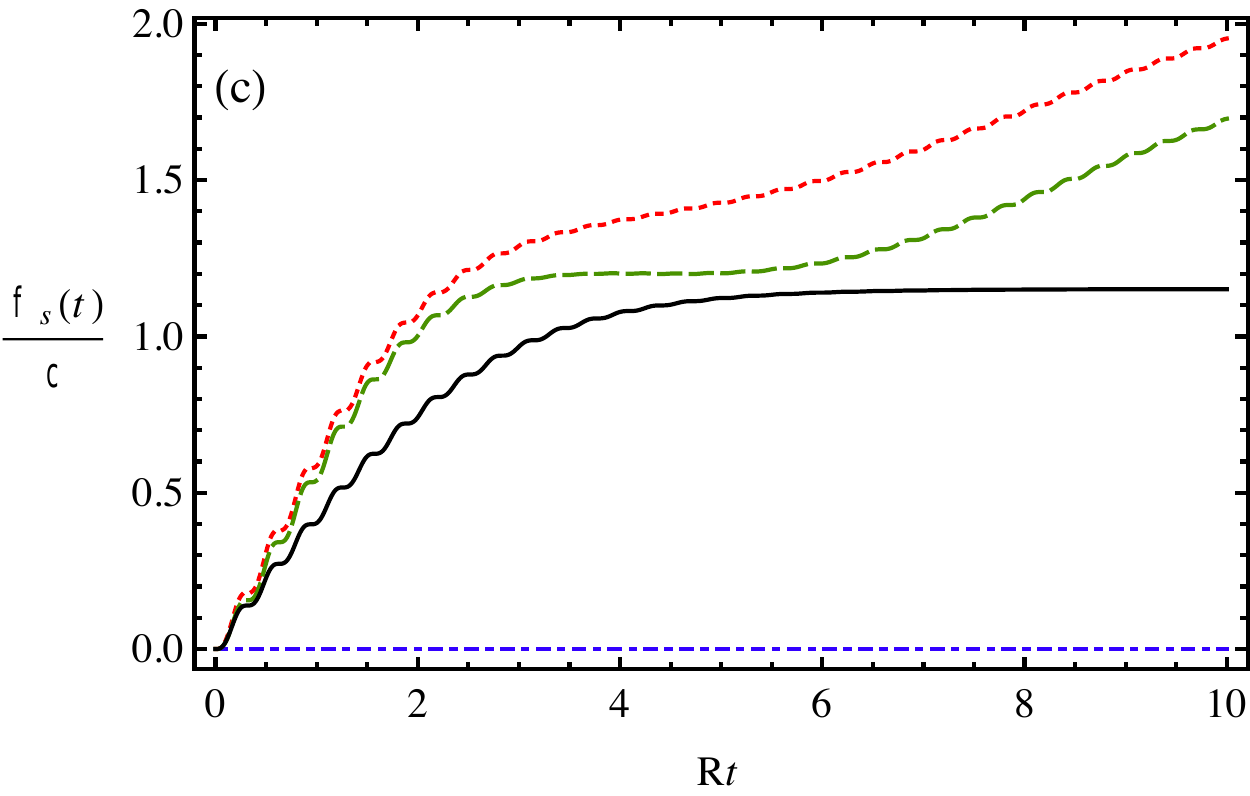,width=8.0cm}
\caption{(Color online) The spatially-averaged signal-mode phase scaled by $\chi$ and plotted versus $Rt$ for the Rabi regime, with $\Delta=0$ and $\Omega_L=10\Gamma D$.The initial conditions and geometric parameters are the same as in figure \ref{Starkns91}.
\label{Rabiphis91}}
\end{figure}

\subsubsection{Conclusions}
\label{SPEconclusions}

In figure \ref{Starkns91}, we see a striking difference between the prediction of the Longitudinal Green's function and both the exact and Paraxial Green's function, with the MWA process being strongly suppressed in the latter case. In Fig. \ref{Rabins91}, we that on-resonance this suppression goes away.
The difference between the off-resonant and on-resonant population dynamics can be attributed to the spatial-variation in the AC Stark-shift of the pump matter-wave which arises due to laser-depletion. The right-circularly polarized light field can be decomposed into a superposition of the bare (undepleted) laser field, and a ``depletion field'', which is opposite in sign, as $\Omega_+(\bfr,t)=\Omega_L+\Omega_d(\bfr,t)$. In the case of a large detuning, the AC Stark-shift of state $|1\ra$ is given by
\beqa
\Delta_{Stark}(\bfr,t)&=&-\frac{|\Omega_+(\bfr,t)|^2}{4\Delta}\nn
&=&-\frac{|\Omega_L|^2}{4\Delta}-\frac{1}{4\Delta}\left(\Omega_L^\ast\Omega_d(\bfr,t)+c.c\right)\nn
&+&\frac{|\Omega_d(\bfr,t)|^2}{4\Delta}.
\eeqa
The depletion field is generated by the induced polarization, and therefore scales as $\Omega_d\sim \Omega_L/\Delta$. Thus to second-order in $1/\Delta$, we have
\beq
\Delta_{Stark}(\bfr,t)\approx -\frac{|\Omega_L|^2}{4\Delta}+\frac{|\Omega_L|^2}{8\Delta^2}(G+G\str)[\psi_1\str\psi_1](\bfr,t).
\eeq
This shows that the spatially-dependent part of the AC Stark-shift scales with the laser-intensity and detuning exactly as the MWA gain term. Thus the effect of pump-depletion is universal, i.e., it does not improve by going further off resonance and increasing the laser intensity. This means that for MWA, {\it there is no such thing as the undepleted-pump approximation in the Stark regime}.  The MWA suppression due to the depletion field should be largest when the real-part of $G$ is strongest, corresponds to the largest $\chi$, and hence the smallest $\cF$, as can be seen in the numerical simulations.

\begin{figure}
\epsfig{figure=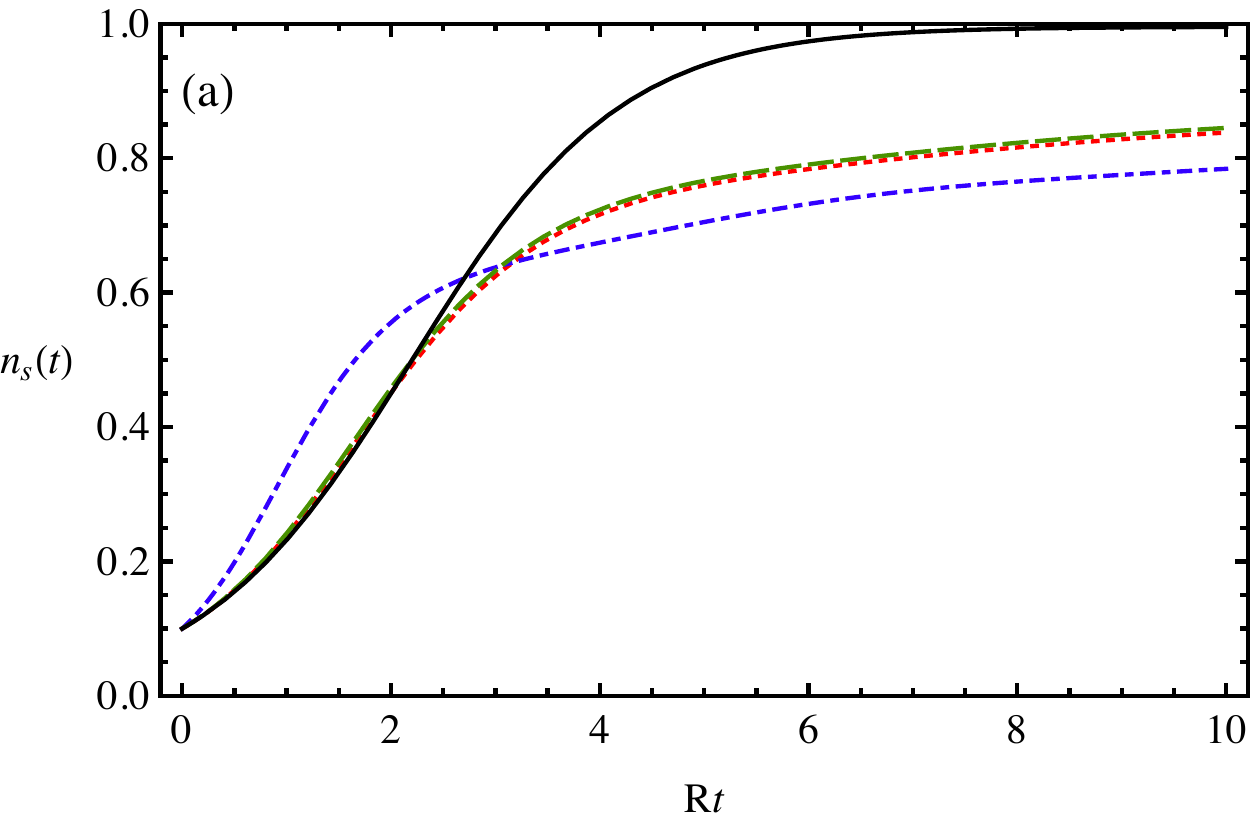,width=8.0cm}
\caption{(Color online) Numerical results for the same parameters as figure \ref{Starkns91}(a), but with the spatially-dependant AC Stark term removed from the field equations, verifying that this removes the MWA supression.
\label{Starkns91removed}}
\end{figure}

\begin{figure}
\epsfig{figure=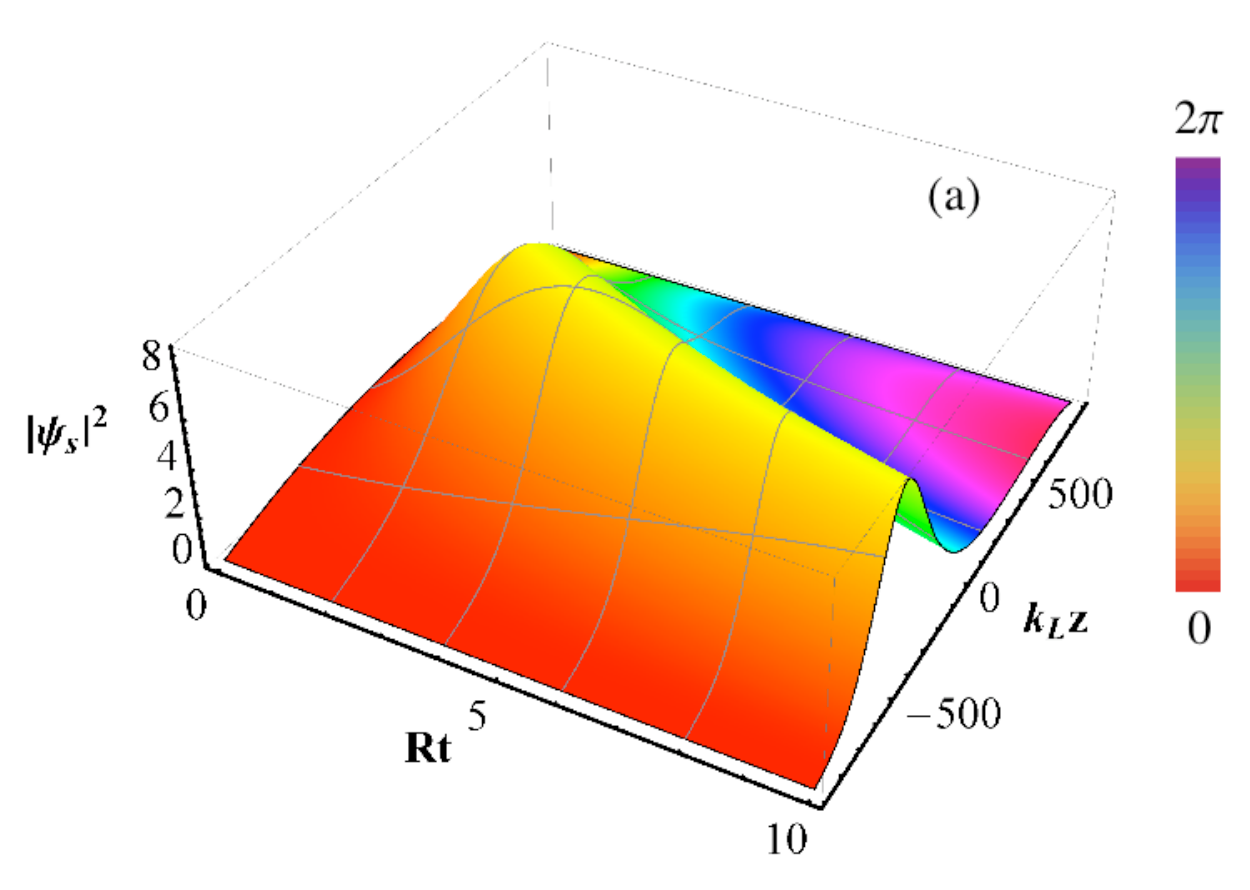,width=8.0cm}
\epsfig{figure=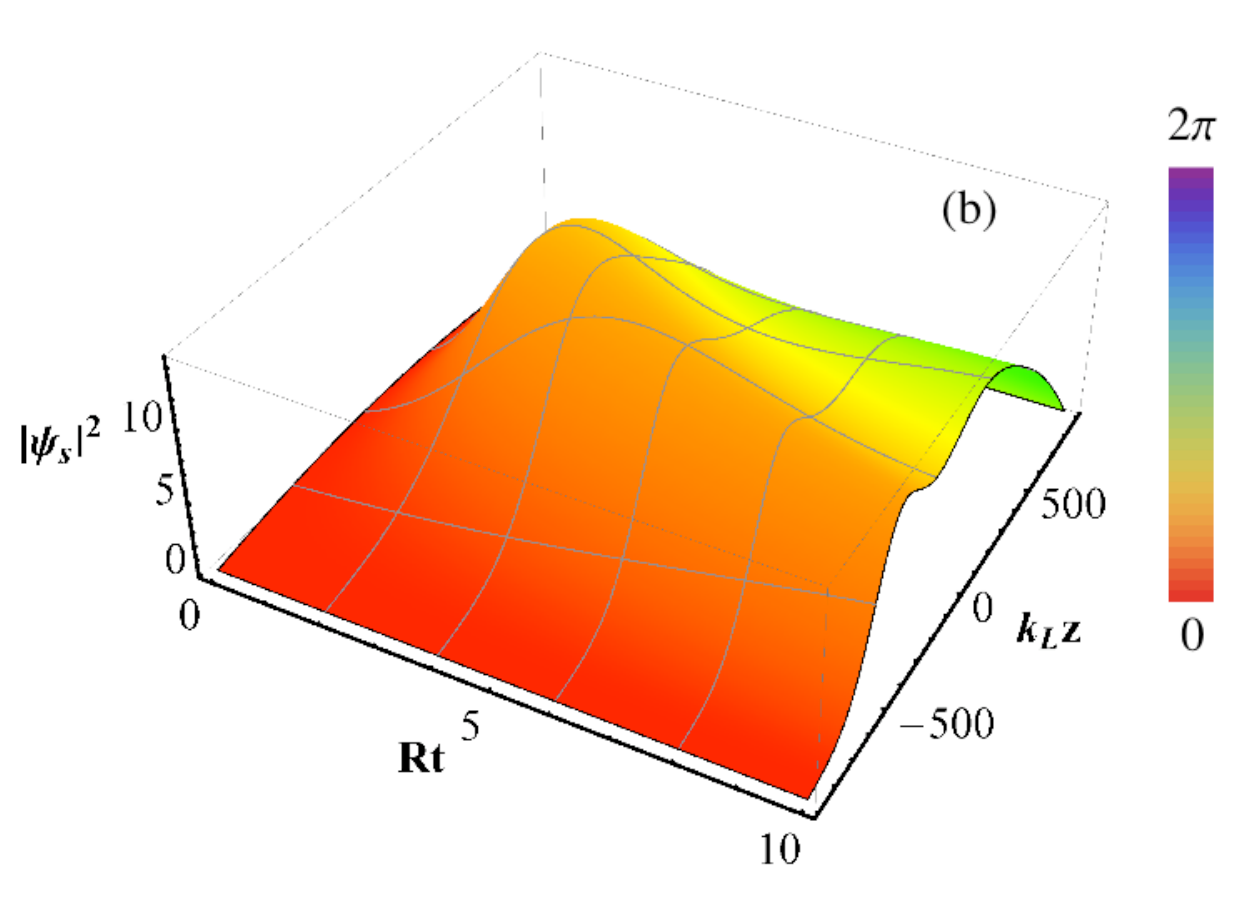,width=8.0cm}
\caption{(Color online) Signal mode density at $\rho=0$ versus $Rt$ and $k_Lz$, for the Stark regime with $\cF=3.18$ and $n_s(0)=0.1$. The shading indicates the phase  of the signal amplitude. In (a) we see that the signal mode is initially amplified at the back of the condensate, but is then reabsorbed, so that signal-mode amplification occurs at the front of the BEC. Note that this effect would not occur when the pump laser is perpendicular to the long-axis. In (b) we show the numerical results with the spatially dependent AC-Stark term removed, showing that reabsorption does not occur, so that the amplification is concentrated towards the back of the condensate.
\label{Stark318phis91}}
\end{figure}

To verify that this explanation is correct, we reproduce Fig. \ref{Starkns91}a, but with the spatially dependent Stark-shift term removed, i.e. the second-term on the r.h.s. of Eq. (\ref{adelimpsi1}). The result is shown in figure \ref{Starkns91removed}, where we see that the effect disappears. To better understand the mechanism for MWA suppression in the Starkj regime, note that 
If the phase differential across the signal-mode is of order $\pi$, then the idler field emitted at the front of the condensate will be reabsorbed rather than amplified. That this is indeed what is going on is shown in Fig. \ref{Stark318phis91}, where we plot the field signal strength $|\psi_s|^2$ and phase, $\arg \psi_s$, at the center of the condensate, $\rho=0$, as a function of $z$ and $t$.
We see that initially the signal mode is built up towards the back of the BEC, as expected, but as the amplification wave moves towards the front, the signal initially built up at the back is reabsorbed. Thus the signal amplification occurs not at the rear, as has been observed experimentally, but at the front of the BEC. The reason this effect was not seen experimentally, is that previous experiments pump from the side, so that laser depletion, and the associated AC Stark modulation does not play a significant role.

Even in the absence of this re-absorption effect, the fully spatial models show a cross-over from a fast-gain process, accurately captured by the QCM model, to a slow-gain process. This cross-over is present in all regimes, and is not related to the spatial AC Stark-shift.  Instead, it is related to the spatial build-up of the idler field, so that atoms at the front of the condensate see a very weak field and are inefficiently transferred from the pump to signal modes, as described by previous authors. Note that even with this effect included, there is still exact agreement with the QCM model at short times. This is because the spatial structure of the idler and pump fields is exactly included in the calculation of $f_R$ and $f_I$, the parameters that govern the QCM model. It is only the spatially-dependent response of the matter-waves that is not included. The cross-over into this slow gain regime is due to the fact that the signal matter-wave is concentrated mainly in the back of the initial condensate region, and the source mode is concentrated in the front region due to depletion at the back. These two modes become highly mis-matched, so that the idler field generated by coherence between them becomes very weak. 

The single-mode QCM model predicted that the dynamics in the Stark and Rabi regimes is universal, with the only difference being in the underlying time-scale of the dynamics. In the full multi-mode model, we see that in addition to a faster dynamics, the on-resonant Rabi regime has significantly better performance due to the absence of the spatially dependent AC Stark-shift of the laser-depletion field. For all regimes, we see that the signal-mode phase-shift scales as $\chi$, which is purely a function of the Fresnel number $\cF$. By decreasing the Fresnel number, one can increase the MWA transfer rate, at the expense of increasing also the non-linear phase-shift. Conversely, by increasing the Fresnel number, one can reduce the phase-shift at the cost of reducing the MWA transfer rate. For a given choice of $\cF$, and hence $\chi$, it is always advantageous in terms of time to work in the on-resonant Rabi regime. Furthermore, if one choses to maximize the MWA gain by taking $\cF\approx 3.18$, then the Rabi-regime has the additional advantage of the absence of idler reabsorption due to the AC Stark-shift of the laser depletion field.

\subsection{Gain versus losses}
Our numerical simulations include atom losses, i.e. decay of the atomic mean-field intensity, which occur due to incoherent scattering of the laser photons into modes other than the idler mode. In this process, a recoiling atom is entangled with a scattered photon, so that both the atomic and optical mean-fields are decreased. This process is suppressed relative to idler emission by a factor of the optical depth $D= f_RN$, due to the collective enhancement of the latter process. Thus losses are minimized by maximizing $D$, which occurs at $\cF=3.18$. 

\begin{figure}
\epsfig{figure=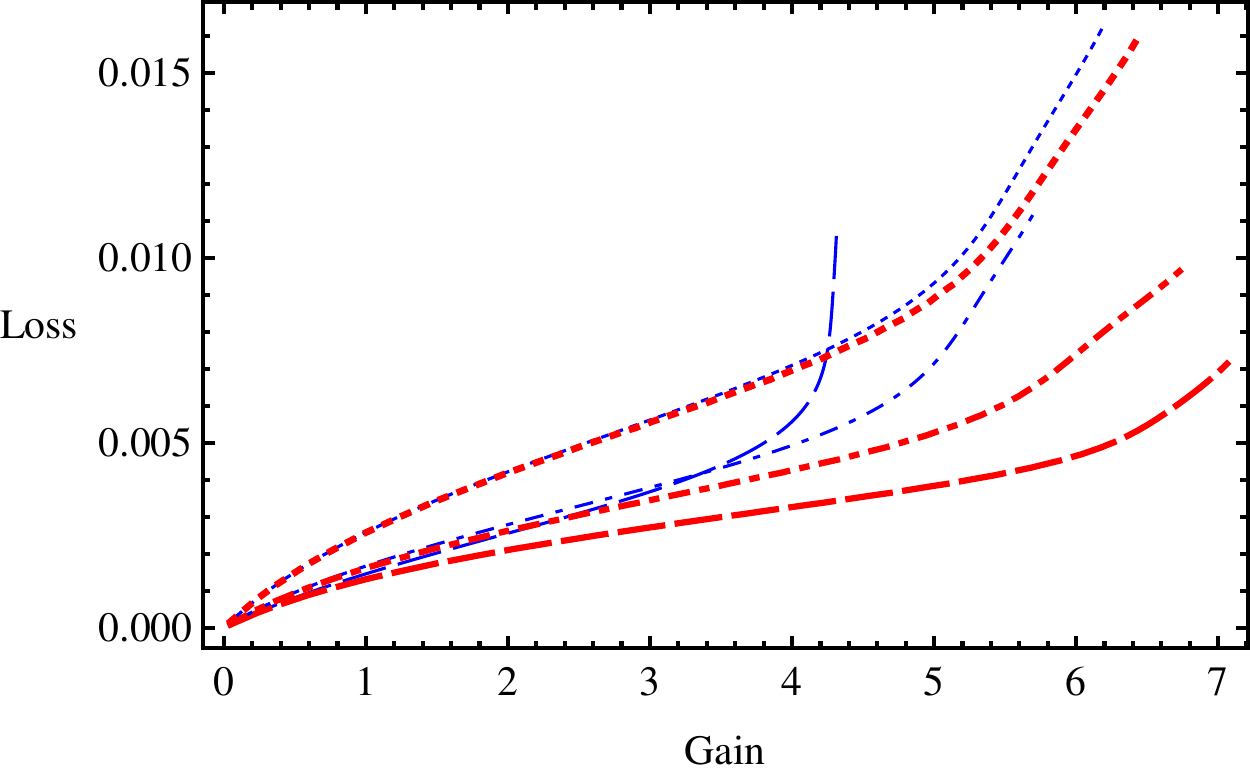,width=8.0cm}
\caption{(Color online) Here we plot the atom-loss percentage versus the signal-mode gain factor calcuated using the full Green's function. The (red) thick lines correspond to the on-resonant Rabi regime, and the (blue) thin lines to the off-resonant Stark regime. The dashed lines correspond to $\cF=3.18$, the dash-dotted lines to $\cF=10.0$, and the dotted lines to $\cF=30.0$, we see that the lowest loss-to-gain ratio occurs on-resonance, and under conditions of maximum gain, $\cF=3.18$.
\label{gainvsloss}}
\end{figure}

One characterization of the efficiency of the MWA process is the gain-to-loss ratio. In Fig. (\ref{gainvsloss}) we plot the gain versus loss trajectory of the signal mode for both the on-resonant Rabi regime (thick red lines), and the off-resonant Stark regime (thin blue lines). The atom loss percentage is 
defined as $\mbox{Loss}=n_s(t)+n_1(t)+n_e(t)$, and the gain factor is defined by $\mbox{Gain}=(n_s(t)-n_s(0))/n_s(0)$.
The dashed lines correspond to $\cF=3.18$, the dot-dashed lines to $\cF=10$, and the dotted lines to $\cF=30$. We see that by this metric, resonant MWA with $\cF=3.18$ is optimal. It is fortuitous that this also corresponds to the case where the MWA transfer process occurs on the fastest time-scale.

\subsection{State transfer dynamics}
Another potential application of the MWA system is the transfer of a quantum state from a matter-wave field onto an optical field. For example, the relative phase between the left and right modes of a double-well BEC could be measured by mapping the relative phase onto an idler field, which is then measured via homodyne detection with the pump laser. This would be accomplished by first removing the double-well potential barrier, so that the two condensates move towards and through each other, and at the moment in which the two modes overlap, a single pump laser is turned on to amplify on of the modes at the expense of the other. The phase of the idler field generated by this process will be given by the relative phase between the two condensates.  In mean-field theory, we cannot make claims regarding squeezed states, however, we believe that conservation laws will be sufficient to guarantee that number- and phase-squeezed states of the two-mode condensate can be mapped onto the idler pulse without destroying the squeezing. 

\begin{figure}
\epsfig{figure=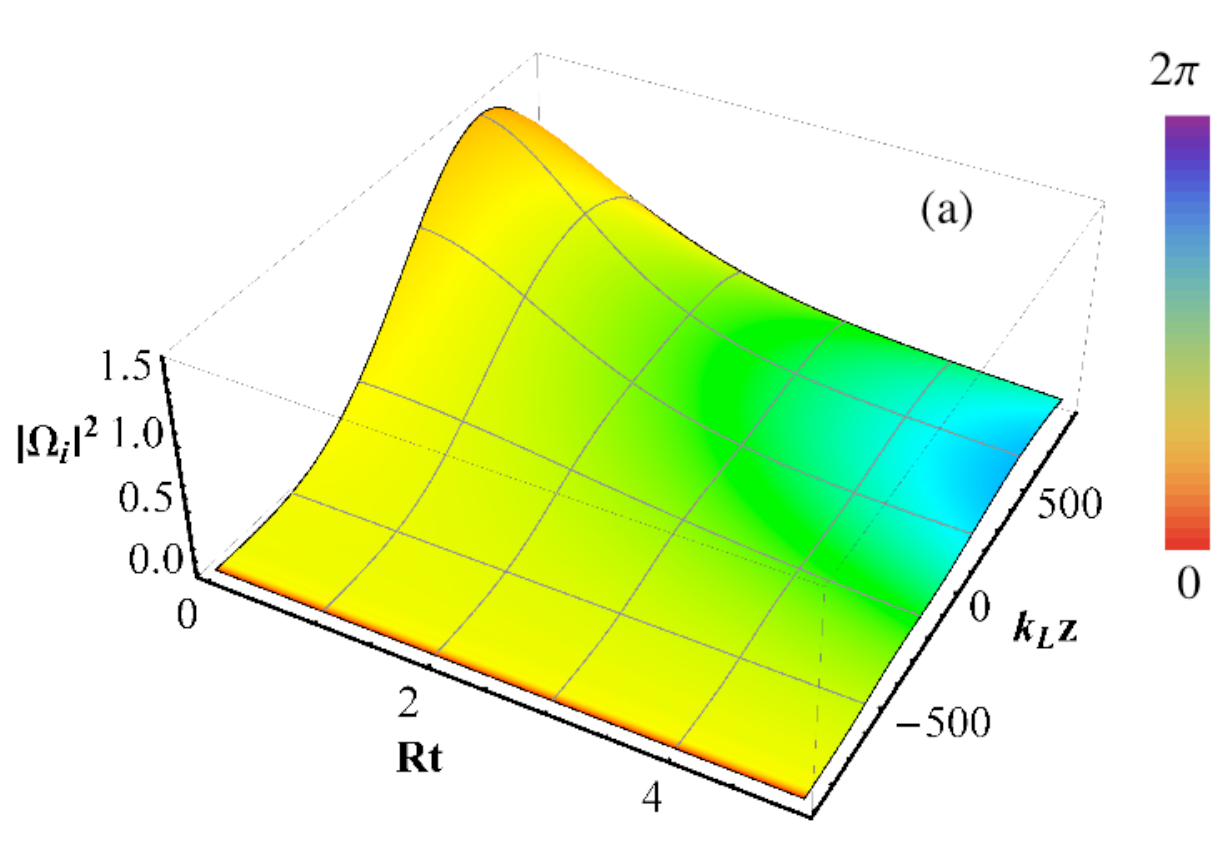,width=8.0cm}
\epsfig{figure=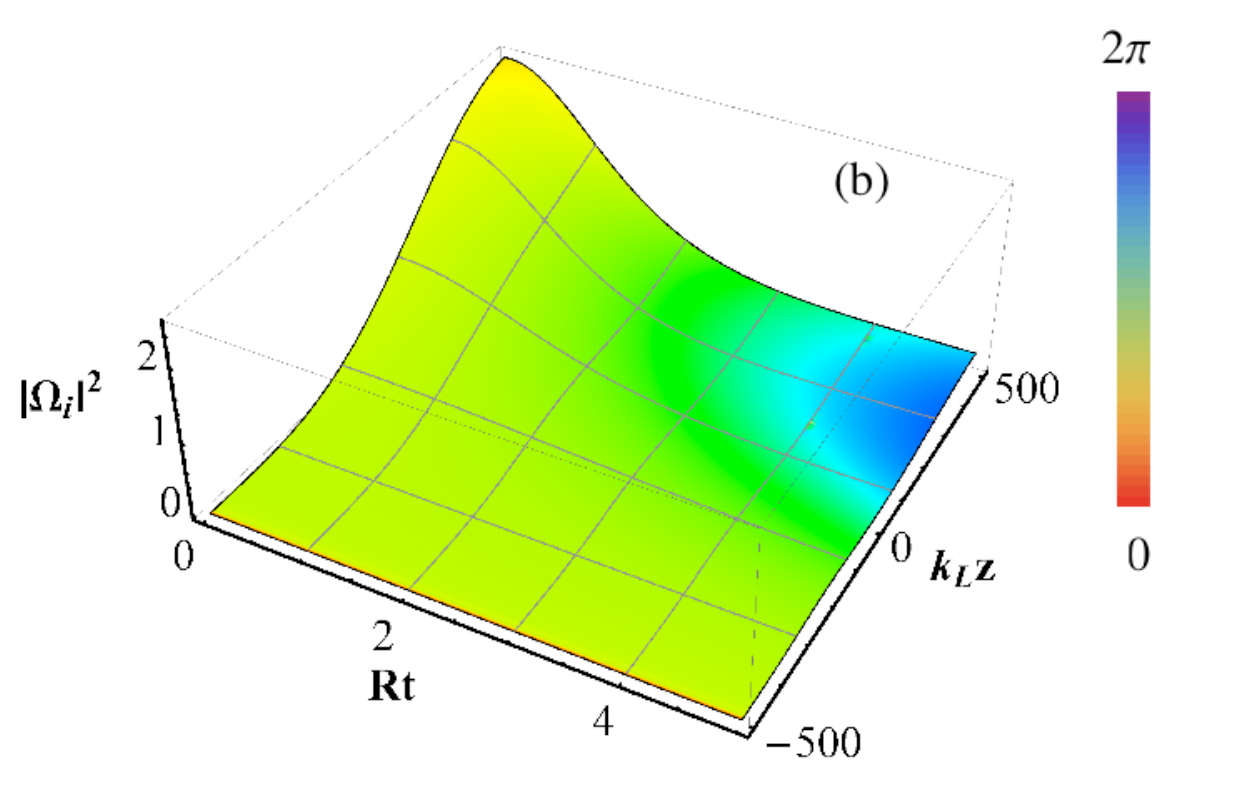,width=8.0cm}
\epsfig{figure=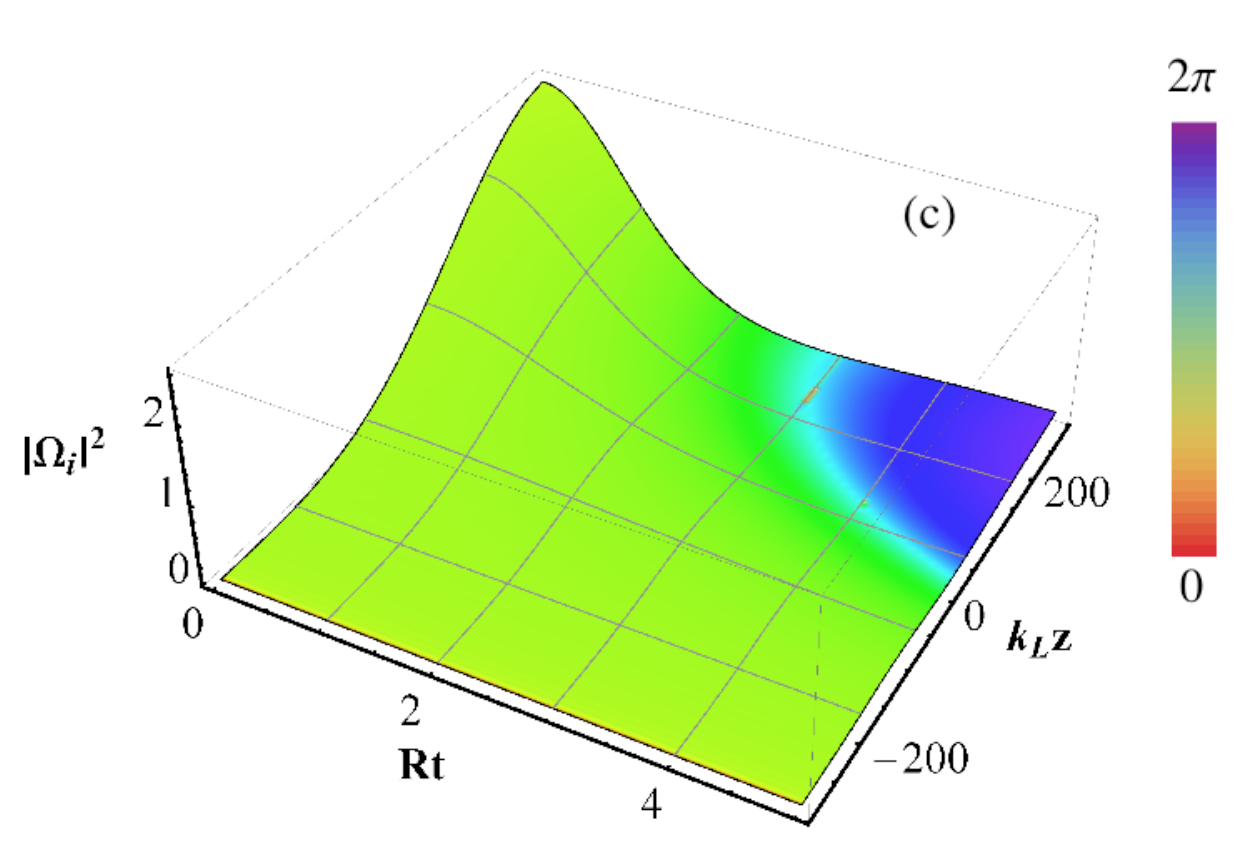,width=8.0cm}
\caption{(Color online) Idler field intensity at $\rho=0$ versus $Rt$ and $k_Lz$ for the off-resonant Stark regime, for the initial conditions $n_s(0)=n_1(0)=0.5$, corresponding to a matter-to-light state transfer protocol. Figures (a), (b), and (c) correspond to $\cF=3.18$, $10.0$, and $30.0$, respectively.
\label{starkSTD}}
\end{figure}

\begin{figure}
\epsfig{figure=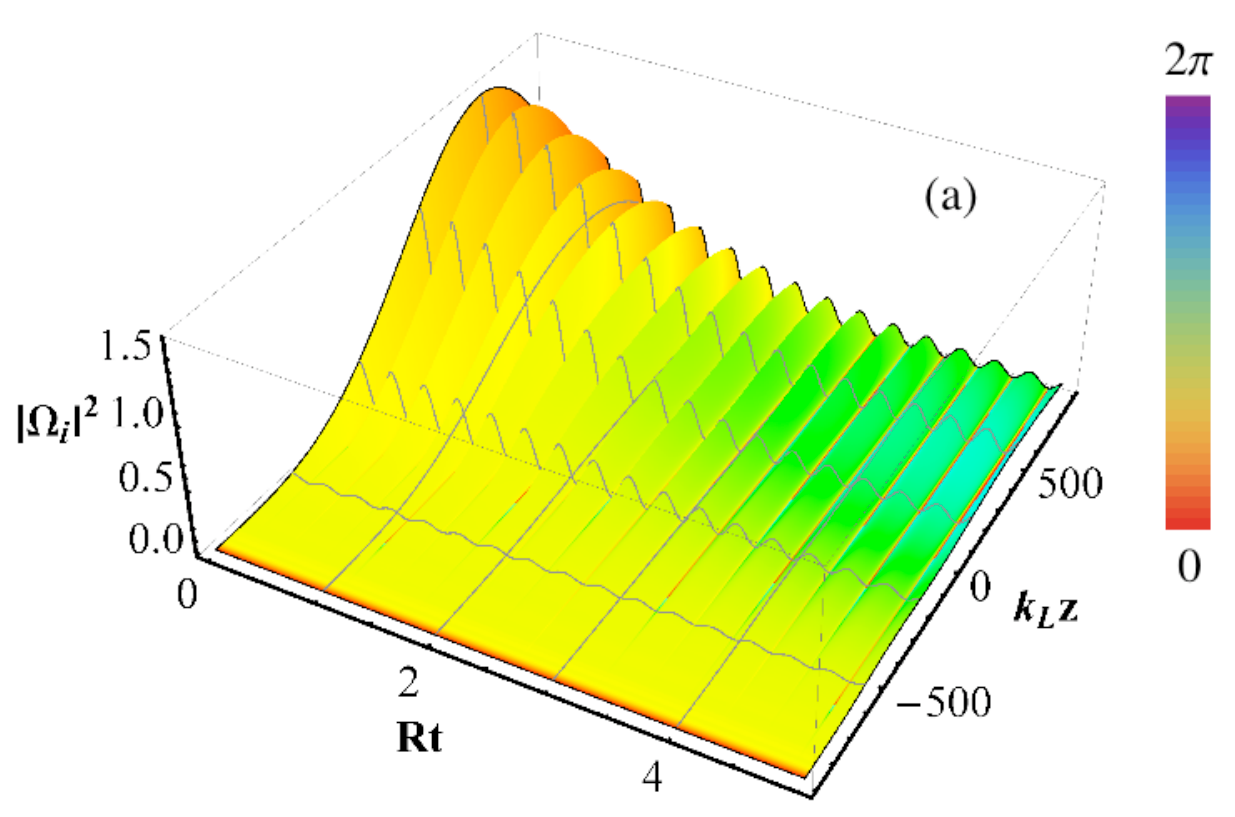,width=8.0cm}
\epsfig{figure=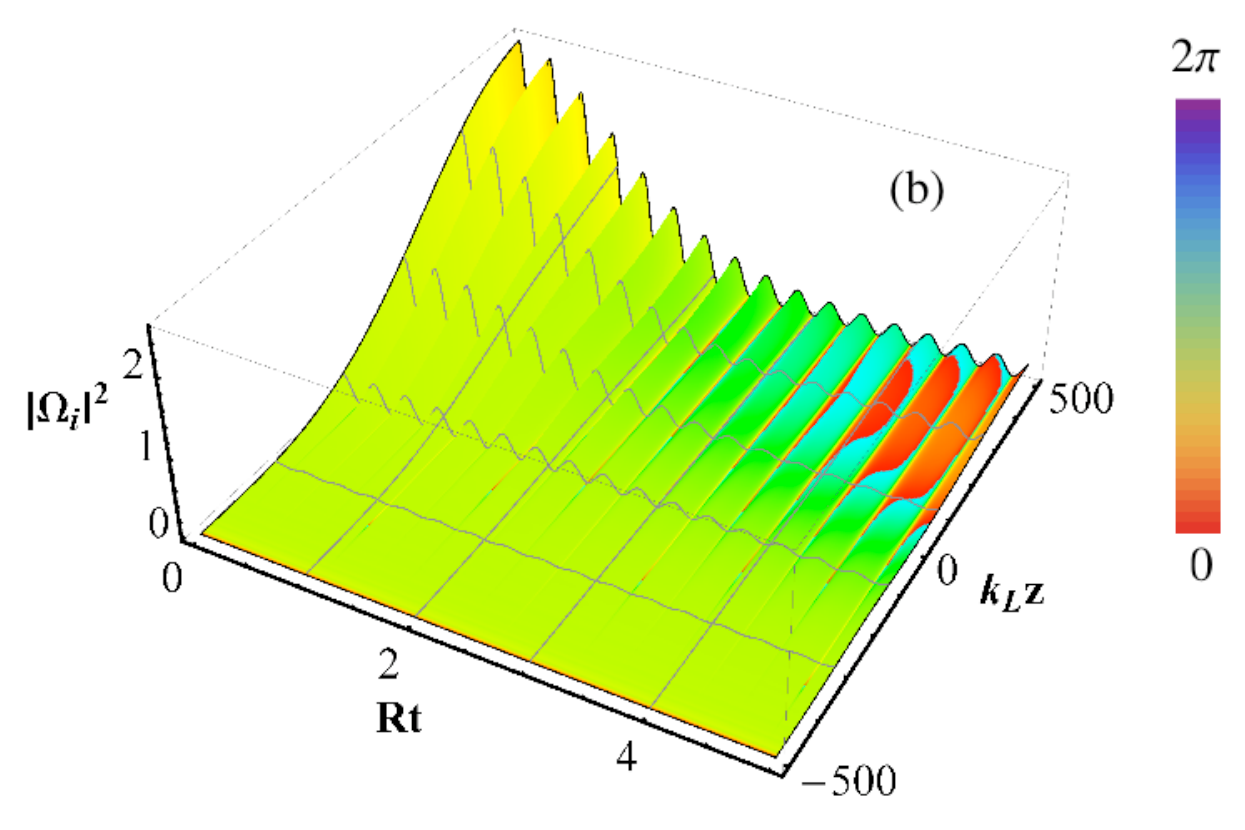,width=8.0cm}
\epsfig{figure=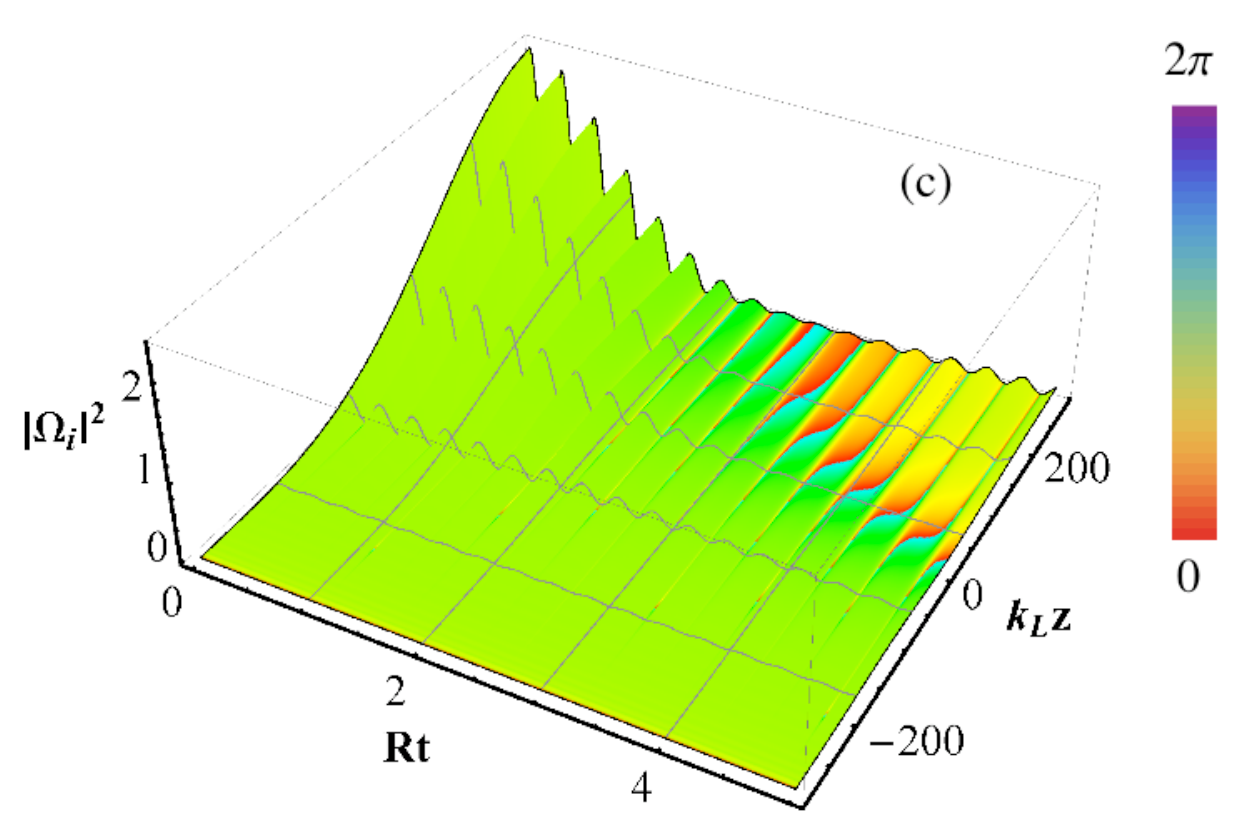,width=8.0cm}
\caption{(Color online)  Idler field intensity at $\rho=0$ versus $Rt$ and $k_Lz$ for the on-resonant Rabi regime, with $\Delta=0$ and $\Omega_L=10\Gamma D$, and for the initial conditions $n_s(0)=n_1(0)=0.5$, corresponding to a matter-to-light state transfer protocol. Figures (a), (b), and (c) correspond to $\cF=3.18$, $10.0$, and $30.0$, respectively.
\label{rabiSTD}}
\end{figure}

To simulate this process, we again perform numerical simulations, but now with initial conditions $n_s(0)=n_1(0)=0.5$. As was seen in the previous section, we find that complete transfer does not occur, due to the fact that atom transfer is inefficient towards the front of the BEC, where the idler field is very weak. Even for a $50\%$ transfer, the number of photons scattered into the idler beam will be of order $N/4$, so that the shot-noise of the idler pulse will be very close to that of the initial par of condensates. In Figures \ref{starkSTD} and \ref{rabiSTD} we plot the idler field 
at the center of the condensate, $\bfr_\perp=0$, versus $R t$ along the x-axis and $k_Lz$ along the y-axis. The height of the surface corresponds to the idler field intensity and the color corresponds to its phase.

In Figure \ref{starkSTD} we show the results for the far off-resonant Stark regime, where the idler field is defined as
\beq
\Omega_i(\bfr_\perp,z,t)=G[\tilde\psi_s\str\tilde\psi_1](\bfr_\perp,z,t).
\eeq
Figures (a), (b), and (c) correspond to the Fresnel numbers $\cF=3.18$, $10$, and $30$, respectively. For $\cF=3.18$, we see a small variation in the idler phase over space and time, which can be attributed to the spatial variation in the AC-Stark shift of the source matter-wave, as described in Sec. \ref{SPEconclusions}. That this effect becomes less visible as $\cF$ is increased is due to the fact that this phase-shift is proportional to $\chi$, which decreases with increasing $\cF$.  The second interesting thing that we see in the plot is that at long times, there is a relatively weak idler field inside the condensate that does not propagate out of the condensate, but is instead re-absorbed at roughly the mid-point, $z=0$. This is the idler field during the second, much slower stage of the MWA process, where the signal and source condensates have evolved into spatially mismatched modes. Note that the region at large $t$ and $z$ where the idler phase increases towards $\pi$ corresponds to a region where the idler amplitude is very small.

In Figure \ref{rabiSTD} we show the results for the on-resonant Rabi regime, where the idler field is defined as
\beq
\Omega_i(\bfr_\perp,z,t)=G[\tilde\psi_s\str\tilde\psi_e](\bfr_\perp,z,t).
\eeq
Figures (a), (b), and (c) correspond to the Fresnel numbers $\cF=3.18$, $10$, and $30$, respectively. Here the phase is taken modulo $\pi$, so that the discontinuous phase slips that occur during each Rabi oscillation are suppressed. We see that the signature of the Rabi oscillations is a modulation of the idler amplitude at the Rabi-frequency of the driving laser. Aside from this modulation, we see that the spatial and temporal uniformity of the idler phase increases as the Fresnel number increases, as with the off-resonant case.

\section{Conclusions}
\label{ds}
Starting from the standard Markovian reservoir theory, we have derived a Green's function approach to the coherent mixing of matter- and light- waves, suitable for modelling the propagation of electric fields inside Bose-Einstein condensates. We have found that at the mean-field level, this approach reproduces exactly the field of an ensemble of ideal radiating dipoles, and is exactly equivalent to solving Maxwell's wave equation. We have studied two approximations to the full Green's function, (1) the paraxial wave equation, that includes diffraction effects, and (2) the longitudinal wave equation, that neglects the transverse component of the Laplacian. Whereas (2) has been used by previous authors, we find that it is inadequate for the task. It does not include the nonlinear phase-shifts associated with the dipole-dipole interaction, which manifest themselves as the real-part of the Green's function. It also does not accurately model the superradiant gain when the length of the condensate length becomes comparable to the diffraction length. The paraxial approximation, on the other hand, can give accurate results provided the aspect ratio is much larger than unity. 

To demonstrate the utility of this approach, we have extended the studies of matter-wave amplification into the
resonant driving regime. We have considered a copropagating Raman MWA scheme and compared the off-resonant Stark regime to the on-resonant Rabi regime. We find that in the off-resonance regime, MWA is suppressed by a spatially dependent AC Stark shift associated with the laser depletion field. As a consequence, MWA is most efficient in the on resonance regime.  We have verified that the rate of mean-field atom losses due to spontaneous emission (heating) scales with detuning and laser intensity in the same way as the MWA gain, so that the gain-to-loss ratio is the same both on and off resonance, and determined solely by the optical depth of the condensate.

In addition, we have compared the single-mode approximation for the matter-waves to the fully spatial propagation, and
 find that the single-mode model gives exact agreement for the short-time evolution. The MWA gain is determed by the optical depth of the condensate. The optical depth $D$, defined in the short time regime, can therefore be computed analytically. We have found that upon scaling by $\sqrt{v}$, where $v=k_L^3V$, with $V$ being the volume, that for sufficiently  large $v$ the gain exhibits a universal behavior which depends only on the Fresnel number $\cF=k_LW^2/L$, which is the ratio of the diffraction length, $L_d=k_LW^2$, to the BEC length, $L$. For $L\ll L_d$, we obtain the usual formula $D\sim n\lambda^2 L$, but for the case $L\gg L_d$, it saturates instead at $D\sim n\lambda^2 L_d$. This leads to a maximum optical depth of $D_{max}\sim N/\sqrt{v}$, occuring at $L\sim L_d$, or more precisely, $\cF=3.18$. In addition to the gain, there is also an associated MWA phase shift, which scales as $\chi$, the ratio of the imaginary collectivity parameter to the real-part. Thus one can generalize the optical depth to a complex parameter $D_{complex}=D(1-i\chi)$.

To summarize, we propose that it is optimal to use resonant driving, instead of the
commonly used off-resonance driving, to perform Raman MWA for the purposes of
amplifying a matter wave and/or mapping quantum states between
optical and matter waves. The reason being that the gain-to-noise ratio is slightly higher and the AC-Stark shift vanishes. This requires a short ($\sim 1$ns) pulse with a Rabi 
frequency satisfying $\Omega_L\ge \Gamma D\sim 10^{10}$.

\end{document}